\documentclass[fleqn,usenatbib]{mnras}

\usepackage{newtxtext,newtxmath}

\usepackage[T1]{fontenc}
\usepackage{ae,aecompl}
\usepackage[normalem]{ulem}

\usepackage{graphicx}	
\usepackage{amsmath}	
\usepackage{amssymb}	
\usepackage{tikz}
\usepackage{dirtytalk}

\title[Stellar streams in Gaia DR2]{Identifying Stellar Streams in \textit{Gaia} DR2 with Data Mining Techniques}

\author[N. W. Borsato et al.]{
Nicholas W. Borsato$^{1}$\thanks{Email: n.borsato@unswalumni.com}, Sarah L. Martell$^{1,2}$, and Jeffrey D. Simpson$^{1}$
\\
$^{1}$School of Physics, UNSW Sydney, Sydney NSW 2052, Australia\\
$^{2}$ARC Centre of Excellence for All Sky Astrophysics in Three Dimensions (ASTRO-3D), Australia\\
}

\date{Accepted XXX. Received YYY; in original form ZZZ}

\pubyear{2019}

\begin{document}
\label{firstpage}
\pagerange{\pageref{firstpage}--\pageref{lastpage}}
\maketitle

\begin{abstract}
Streams of stars from captured dwarf galaxies and dissolved globular clusters are identifiable through the similarity of their orbital parameters, a fact that remains true long after the streams have dispersed spatially. We calculate the integrals of motion for 31,234 stars, to a distance of 4 kpc from the Sun, which have full and accurate 6D phase space positions in the \textit{Gaia} DR2 catalogue. We then apply a novel combination of data mining, numerical and statistical techniques to search for stellar streams. This process returns five high-confidence streams (including one which was previously undiscovered), all of which display tight clustering in the integral of motion space. Colour-magnitude diagrams indicate that these streams are relatively simple, old, metal-poor populations. One of these resolved streams shares very similar kinematics and metallicity characteristics with the Gaia-Enceladus dwarf galaxy remnant, but with a slightly younger age. The success of this project demonstrates the usefulness of data mining techniques in exploring large data sets.
\end{abstract}




\section{Introduction}
Stellar streams are the footprints of our Galaxy's evolution. Their creation is a by-product of two possible Galactic processes: the end-result of the tidal disruption of globular clusters, or the accretion of dwarf galaxies \citep{2018_example_price_wheel,Helmi_a}. Locating and classifying both varieties of streams is, therefore, a means to uncover our Galaxy's past \citep{johnson_soderblom_1987}. Furthermore, the discovery of streams offers several additional benefits to the understanding of Galactic processes. The total number of streams can in principle place a lower limit on past accretion events in the Galactic halo \citep{Bullock_Johnston_2010}, while those created as a result of gravitational disruption can offer insight in the formation and evolution of globular clusters \citep{Balbinot_2018, Bose_2018}. Streams can also act as probes into the Milky Way dark matter halo, providing the means to map its mass distribution and shape \citep{Johnston_1996, Ibata_2001, Koposov_2010, Law_and_Maj_2010, Bowden_2015, Bovy_2016, Malhan_3,Malhan_Ibata_2019}. 

Stellar streams are found using a variety of different search methods. A common theme amongst them is to find clusters of measured stellar parameters in astronomical surveys. Clusters often indicate stars of the same origin. Options range from abundance-, kinematic-, and location-based techniques, or combinations of all three. 

Streams formed from disrupted globular clusters maintain their chemical similarity, offering a means to detect them if their populations can be resolved from the stellar background with sufficient contrast. Matched Filter techniques incorporate the colour-magnitude weighting of stars to find structures that belong to the same origin \citep{Grillmair_Freeman_1995,Rockosi_2002, Balbinot_2017}. An abundance-based technique, however, is reliant on spectroscopic data to separate stream members from the background stars. If data quality is poor, it can lead to missed detections \citep{Malhan_2}. Additionally, detection can fail if there is significant dispersion in the abundance patterns of the stream. Notable streams that have been identified using the matched filter technique include the Palomar 5 tidal tails \citep{Odenkirchen_2001}, GD-1 \citep{Grillmair_and_Dinotos_2006}, Orphan \citep{Belokurov_2006}, Eridanus, Palomar 15 \citep{Myeong_2017}, and the streams detected in the Dark Energy Survey \citep{Shipp_2018}. 

Finding clusters of stars within kinematic spaces can also lead to the detection of streams. Conventional techniques rely on either space-velocities to locate co-moving groups of stars or identifying clusters in a data set's \say{integrals of motion} values. Integrals of motion refer to the angular momentum actions defined by \cite{binney_tremaine_2008} and total orbital energy of a star. These quantities are ideal for stream searches as they are argued to be conserved, or quasi conserved \citep{Helmi_a, Helmi_b, Myeong_2, Koppel, Meingast_2019}. Both techniques depend on locating clustered sets of kinematic parameters and proving these clusters to be statistically significant. 

Successful stream searches using velocity parameters have resulted in the detection of the Aquarius, the VelHel-1 to -9 streams, and the S1-4, C1-2, Rg1-7 and Cand8-17 sub-structures \citep{Williams_Galaxia, Helmi_2017, Myeong_1}. While difficulties can arise if there is a strong velocity gradient along the stream, the technique is viable for detecting streams which have not undergone a significant degree of phase mixing. Phase mixing is a collective name for kinetic energy exchanges between the stream stars and other constituents of the Galaxy \citep{Klement_2010}. 

If phase mixing is an issue, integrals of motion can be relied upon instead of velocities to search for streams kinematically. Relying on conserved quantities has led to the discoveries of the H99, $S_1, S_2$ and $S_3$ streams \citep{Helmi_a, Klement_2009} and the rediscovery of S1-4, C1-2, Rg1-7 and Cand8-17 in a follow-up study provided by \cite{Myeong_2}. Integrals of motion provide robust information for stream searches because they are conserved over long periods. However, they require knowledge of the six-dimensional phase space positions of each star \citep{Malhan_2}. This information is usually difficult to obtain for a large population of stars and is very challenging to acquire for very distant stars. This impediment limits the scope of integrals of motion-based studies to the local Galaxy. Nevertheless, many studies use the integrals of motion to search for streams. For example, in \textit{Gaia} DR2 \cite{Koppel} searched halo stars within 1 kpc of the Sun, and found 5 potential stream candidates with clearly identifiable clustering in their integrals of motion.

Attempts to mitigate the weaknesses in each of these approaches usually involve combining methods. The Cetus Polar Stream \citep{Newberg_2009} was detected using both metallicity and velocity information to differentiate the stream from the tidal tail of the Sagittarius Dwarf galaxy \citep{Law_2005}. The recently developed STREAMFINDER algorithm \citep{Malhan_1} combines aspects of spatial, chemical and kinematic techniques in conjunction with probabilistic arguments to search for stellar streams and characterise their shapes and their distribution in the Galaxy. STREAMFINDER has had a great deal of success in locating previously unknown streams within the \textit{Gaia} database, naming these streams after great lakes in Greek and Norse mythology \citep{Malhan_4, Ibata_2019_Malhan_5}. Combined search attempts are ideal if the required information is available since an agreement between multiple independent techniques adds to the legitimacy of any detection made.

With the continual increase in the size of wide-field astrometric, imaging and spectroscopic surveys \citep{LAMOST,SDSS,RAVE,RAVEon,GALAH_DR2,Gaia_B}, the necessity of using algorithmic search methods is becoming more and more apparent \citep{kmeans,friend-of-friends}. Data mining algorithms that search for clusters in data sets are well suited for this task as they can handle more challenging and higher-dimensional problems objectively, allowing one to focus primarily on verifying and characterising the properties of the discovered streams. In this study, we use the clustering algorithm DBSCAN \citep{Ester_1996} to search for clustering in the integral of motion space using stars with full 6D phase space information in the \textit{Gaia} DR2 data set \citep{Gaia_B}, and analyse the colour-magnitude diagrams of these candidate streams. Our goal is to replicate the results of \citet{Koppel, Koppelman_2018_2} while locating additional stream candidates, using a more stringent data quality selection on the \textit{Gaia} data set.

We describe the data set in Section \ref{sec:data}, outlining the subset restrictions and quality criteria we applied. Section \ref{sec:method} explains the process of our stream finding method: \ref{sec:ApplicationDB} outlines how we actually applied the data-miner to the data set, \ref{sec:input} will explain how we chose our search parameters. The determination of statistical significance is explained in \ref{sec:Significance}.  Section \ref{sec:results} presents our results, with our findings broken up into two parts; Section \ref{sec:Comparison Cut Search} presents a comparison of our results to the literature, and our finalised results presented in \ref{sec:Quality Cut Search}. Discussion of the results and interpretation of our overall findings of the paper are presented in Section \ref{sec:discussion}.

\section{The Data Set}
\label{sec:data}
Our search for potential stream candidates use the astrometric and kinematic data available in the second \textit{Gaia} data release \citep{Gaia_B}. In this data set, 7,224,631 stars have full six-dimensional phase space and photometry information. From this subset, we create two data sets. The first, the \say{replica data set}, uses the same selection as \cite{Koppel}, which we will use to compare the results of our method to theirs. The second data set, the \say{restricted data set}, is a subset of the replica data set with stricter selection criteria to help ensure that our results are reliable.

Creating the replica data set involved only accepting stars with a parallax uncertainty of less than 20\%, leaving us with a total of 6,447,952 objects. Restricting the parallax uncertainties to this level enables inversion of the parallax measurement to provide a distance measure. Inverting parallaxes typically creates difficulties when attempting to incorporate uncertainties due to the hyperbolic nature of the conversion. A potential remedy for this issue is to use Bayesian distance priors to calculate the uncertainties; however, inverting parallaxes with percentage errors less than 20\% is an acceptable alternative \citep{Parrallax-cut-off}. We use this replica data set to compare our results to \cite{Koppel}.

For the restricted data set, we apply a correction to the \textit{Gaia} data set, and then apply an additional data quality selection. We correct for the global \textit{Gaia} parallax zero-point offset by applying a $+0.052$ mas correction to the parallax measurements \citep{Zinn_Paralaxx_Offset,Leung_Bovy_Parallax_Offset} which did not change the overall number of stars in the data set. The quality selection cut only accepted stars with a minimum number of radial velocity transits \textsc{$rv\_nb\_transits$} $\geq 5$ \citep{Marchetti}. This cut reduced the data set to 4,533,363 objects red, and was the only additional cut that we applied.

We convert the equatorial sky coordinates, proper motions, parallax and
radial velocity measurements from \textit{Gaia} to Cartesian Galactocentric coordinates using \textsc{astropy} \citep{Astropy}. In this system, the Sun lies on the X-axis of the Cartesian plane at a distance of 8.3 kpc from the origin \citep{Dist_From_Gal_Center} and 27 pc above the Galactic midplane \citep{Height_From_Galc_Cen}. The Z-axis projects towards the Galactic north pole, and the Y-axis aligns with the direction of Galactic rotation. For velocity parameters we use the convention adopted by \cite{johnson_soderblom_1987} with the Cartesian representation of the motions of stars defined as (U,V,W) where U is the direction moving radially toward the Galactic centre, V is the circular tangential velocity in the direction of Galactic rotation, and W is the velocity directed towards the Galactic North Pole. We correct for solar motion using the peculiar velocity correction values of $(U, V, W)_{\odot} = (11.1^{+0.69}_{-0.75}, 12.24^{+0.47}_{-0.47}, 7.25^{+0.37}_{-0.36})\,\rm{km s^{-1}}$ supplied by \cite{Schonrich_2010} and \cite{Bovy_2015} to adjust the circular velocity and Solar radius while also placing all stars in the Local Standard of Rest \citep{McMillan_LSR} $\text{V}_{\text{LSR}} = 232\;\rm{km s^{-1}}$.

The Z component of angular momentum $L_{Z}$, perpendicular angular momentum $L_{\perp}$\footnote{$L_{\perp}$ is calculated by adding the X and Y components of angular momentum together in quadrature: $L_{\perp}^{2} = L_{X}^2 + L_{Y}^2$}, and total energy $E$ values are calculated using the library package \textsc{gala} \citep{Gala}. To model the gravitational potential of the Galaxy, we use the potential class MilkyWayPotential in \textsc{gala}. The constituents of this potential consist of the following analytic models: a Hernquist potential to model the bulge and nucleus of the Galaxy \citep{hernquist_1990}, a Miyamoto-Nagai \citep{Miyamoto_pot} potential to shape the Galactic disk \citep[taken from][]{Bovy_2015}, and a standard Navarro-Frenk-White potential \citep{Navarro_Potenial} to model the halo. Using the phase-space coordinates and the specified gravitational potential, we can calculate the integrals of motion for each star in the data set.

Following previous stream search publications, we focus on the halo of the Galaxy \citep{Helmi_b, Klement_2009, Williams_Galaxia, Bernard_2014, Koppel, Ibata_2019_Malhan_5}. We choose halo stars in \textit{Gaia} DR2 by selecting stars with large total velocities relative to the Local Standard of Rest \citep{Nissen_2010, Koppel}. Using a selection of $|\text{V}-\text{V}_{\text{LSR}}|>210\;\rm{km s^{-1}}$ results in a reduced subset of 152,865 stars for the replica data set, and 62,527 stars for the restricted data set. It is important to note that this approach excludes halo stars with disk-like velocities from the search sample \citep{Nissen_2010, Bonaca_2017, Koppel, Koppelman_2018_2} while also resulting in some contamination from thick disk stars which may have large vertical velocities.

\section{Stream Searching With DBSCAN}
\label{sec:method}
Our stream search begins within the local Solar neighbourhood (out to 1 kpc) before extending the sample to include stars across a broader region of the Galaxy. We observe how far any streams detected within the local halo extend before we can no longer confidently identify them. We only search for potential stream targets at a distance $< 4$ kpc, as increasing the range further leads to cumbersome computational times.

A fundamental property of clustered data points is the density contrast between points inside the cluster versus points outside. Density-Based Spatial Clustering of Applications with Noise \citep[DBSCAN,][]{Ester_1996} is a data mining method which utilises this fact to locate clustered data points. The algorithm requires two input parameters: a characteristic distance $\epsilon$ \citep{Ester_1996, Chen_2018} and a minimum cluster member number termed $MinPts$. DBSCAN selects an arbitrary point within the data set and retrieves all other data points reachable within a distance of $\epsilon$. If the number of other data points within $\epsilon$ exceeds $MinPts$, the selected data point is a core-point. DBSCAN then moves from the core-point to another point within the $\epsilon$ range and determines if that point too also satisfies the core-point criteria. If the test is successful, the cluster size expands to include that point. The process then repeats with another point within range. Data values are assigned a border-point label if they lie within a distance of $\epsilon$ of a core-point, but fail to classify as a core-point. A cluster is complete when border-points surround the core-points. At this stage, DBSCAN moves to a data point outside the cluster and repeats the process. Any data point that does not have the specified $MinPts$ number of data points within a distance of $\epsilon$, and no core-points in range is labelled as a noise-point. Every cluster is given a unique integer label, with noise-points taking a label of -1. 

\begin{figure}
    \centering
    \begin{tikzpicture}[scale=1.5]
    \draw [fill] [blue] (1.4967141530112327,-0.13826430117118466) circle  [radius=0.025] ;
    \draw [fill] (1.6476885381006925,1.5230298564080254) circle  [radius=0.025] ;
    \draw [fill] [red] (0.765846625276664,-0.23413695694918055) circle  [radius=0.025] ;
    \draw [fill] (2.5792128155073915,0.7674347291529088) circle  [radius=0.025] ;
    \draw [fill] [blue] (0.5305256140650478,0.5425600435859647) circle  [radius=0.025] ;
    \draw [fill] [red] (1,0) circle  [radius=0.025];
    
    \draw [dashed, ultra thick] [blue] (1.4967141530112327,-0.13826430117118466) circle  [radius=1];
    \draw [dashed] (1.6476885381006925,1.5230298564080254) circle  [radius=1] ;
    \draw [ultra thick] [red] (0.765846625276664,-0.23413695694918055) circle  [radius=1] ;
    \draw [dashed] (2.5792128155073915,0.7674347291529088) circle  [radius=1] ;
    \draw [dashed, ultra thick] [blue] (0.5305256140650478,0.5425600435859647) circle  [radius=1];
    \draw [ultra thick] [red] (1,0) circle  [radius=1];
    
    \draw [fill=red] (3.02,-0.65) rectangle (2.92,-0.75);
    \node [right] at (3,-0.7) {Core Point};
    
    \draw [fill=blue] (3.02,-0.85) rectangle (2.92,-0.95);
    \node [right] at (3,-0.9) {Border Point};
    
    \draw [fill=black] (3.02,-1.05) rectangle (2.92,-1.15);
    \node [right] at (3,-1.1) {Noise Value};
    
    \draw [ultra thick] [->] [red] (1,0) -- (1.57357643635,-0.81915204428);
    \node [scale=1.2][above] at (1.1, -0.7) [black] {\textbf{$\epsilon = 1$}};
    
    \end{tikzpicture}
    \caption{Visual representation of the DBSCAN process. In this example $MinPts=3$ and $\epsilon=1$. Red points are core members that have at least three data values within the $\epsilon$ range. Blue points are border points which are a part of a cluster but have fewer than three members in range. Black points, which are the noise points, are data values that are not in range of any clusters and do not have enough points close enough to form a cluster of their own.}
    \label{fig:DBSCAN_Image}
\end{figure}
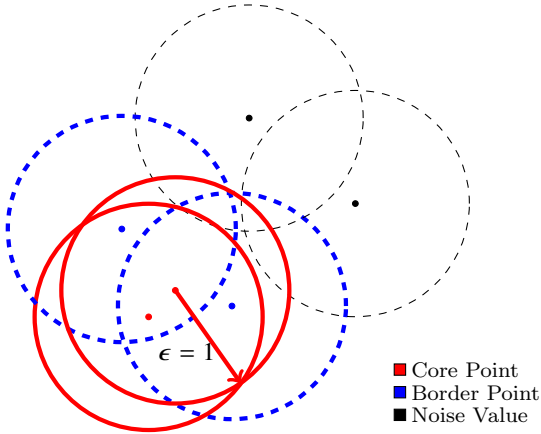

Figure \ref{fig:DBSCAN_Image} offers a visual representation of how DBSCAN operates. In the example, $MinPts = 3$ and $\epsilon =1$. One can see that two points have three data values within a distance of $\epsilon$, and so have been classified as core-points. Two points have fewer than three data values within range and so are classified as border-points while two data points have no neighbours within a distance of $\epsilon$ and are given noise classifications. 

A significant advantage to using DBSCAN is its versatility and applicability to data of arbitrary dimensions. Isolating clusters in higher dimensional space creates a stronger argument that an actual cluster is present, as it is less likely that the apparent structure is coincidental. Additionally, DBSCAN will produce the same cluster patterns every time provided the search parameters remain the same. These aspects are highly useful for stream detection.

\subsection{Applying DBSCAN to the data set}
\label{sec:ApplicationDB}
DBSCAN works under the assumption that all data points are located precisely at their recorded position and draws clusters where the points stand. However, this is not realistic as measured values have uncertainties, which means that DBSCAN may potentially classify clusters that are statistical anomalies created by noise in the data rather than actual clustered sets of stars. Therefore, we perform searches on multiple Monte-Carlo iterations of the search subset to take the measured uncertainties into account.

Positional components, proper motions, parallaxes, and radial velocities were modelled as multivariate Gaussians in a framework similar to \cite{Marchetti} with the joint probability distribution of these variables represented by the following equation:

\begin{equation}
    f(\hat y) = \frac{1}{2\pi^{n/2}}|\Sigma|^{\frac{1}{2}} \;\rm{exp}\Big(\frac{-(\hat y - m)'|\Sigma|^{-1}(\hat y - m)}{2}\Big)
    \label{prob}
\end{equation}

which describes the probabilities with which the random variable $\hat y$ takes on values jointly in its $n$ elements. The variables $m$ and $\Sigma$ refer to the mean vector and covariance matrix of the distribution respectively. The mean vector is given by

\begin{equation}
    m = [\alpha, \delta, \pi, \mu_{\alpha}, \mu_{\delta}, v_{\rm rad}]
\end{equation}

where $\alpha$ and $\delta$ are right ascension and declination, $\pi$ is the parallax, $\mu_{\alpha}$ and $\mu_{\delta}$ are the proper motions in right ascension and declination, and $v_{\rm rad}$ refers to the radial velocity. In addition to this, we constructed a $6\times 6$ covariance matrix of

\begin{equation}
    \Sigma = 
    \begin{pmatrix} 
        \sigma^{2}_{RA} & \dots & \rho(\alpha,v_{\rm rad})\sigma_{RA}\sigma_{v_{\rm rad}} \\
        \vdots & \ddots & \vdots \\
        \rho(\alpha,v_{\rm rad})\sigma_{RA}\sigma_{v_{\rm rad}}
        & \dots & \sigma^2_{v_{\rm rad}}
    \end{pmatrix}
\end{equation}

where $\sigma_{i}$ denotes the standard deviation of each variable in the mean vector, and $\rho(i,j)$ denotes the correlation coefficients of the parameters $i$ and $j$. 

With these parameters we can repeatedly sample from $\hat y$, providing random samples of each star's phase space parameters in line with what we would expect based on the known correlations and uncertainties. We note that radial velocities within the \textit{Gaia} DR2 data set are uncorrelated to the other astrometric parameters. Therefore, correlation coefficients between radial velocity and the astrometric parameters are zero, and the standard deviations are assumed to follow a Gaussian distribution.

Using the covariance matrix, we sample 200 potential integrals of motion values for each star in the search subset. A sample of 200 was large enough to reproduce the results consistently while also minimising the search time. DBSCAN is then run on each of these subsets to produce an initial set of clusters within each random sample. 

Clusters should, in principle, remain coherent even when subjected to the random deviations produced by uncertainties. If any cluster drawn by the algorithm in one Monte-Carlo iteration does not appear in another, the cluster is likely just an artefact produced by the random deviations of the Monte-Carlo sampling. We remove any clusters that do not share at least four stars in common with another cluster in one of the other Monte-Carlo sets, which effectively cleans the data set of statistical anomalies that fail to re-occur when the data is allowed to vary within the errors. This process effectively eliminates clusters that cannot be conclusively proven to exist based on the uncertainties of \textit{Gaia} measurements.

At this point, the clusters found in each Monte-Carlo subset are combined to create an aggregate subset. In this subset, we apply DBSCAN one final time with re-optimised search parameters. This final application of DBSCAN acts to join any partially formed groups together with other partial pieces. After the clusters are merged, they are extracted from the data set and have all duplicates removed. 

\subsection{Specifying Input Parameters}
\label{sec:input}
When choosing our input parameters we set $MinPts = 4$. This specification enables sparsely populated stream candidates to be accepted by the algorithm, while still imposing that some clustering must be present initially. Our method of choosing an $\epsilon$ value relies on the use of a 4-distance plot \citep{Ester_1996}, which is a ranked plot of the standardised distance from each data point to its fourth nearest neighbour. As the magnitudes of the integrals of motion can vary significantly, these quantities need to be standardised. We do this through normalising each of the integrals of motion individually, taking the mean to be zero and scaling to unit variance. Normalisation is achieved using the StandardScaler class in the \textsc{scikit-learn} library \citep{scikit-learn}. Extracted $\epsilon$ values are measured in this standardised distance, and are unitless.

\begin{figure}
    \centering
    \includegraphics[scale=0.22]{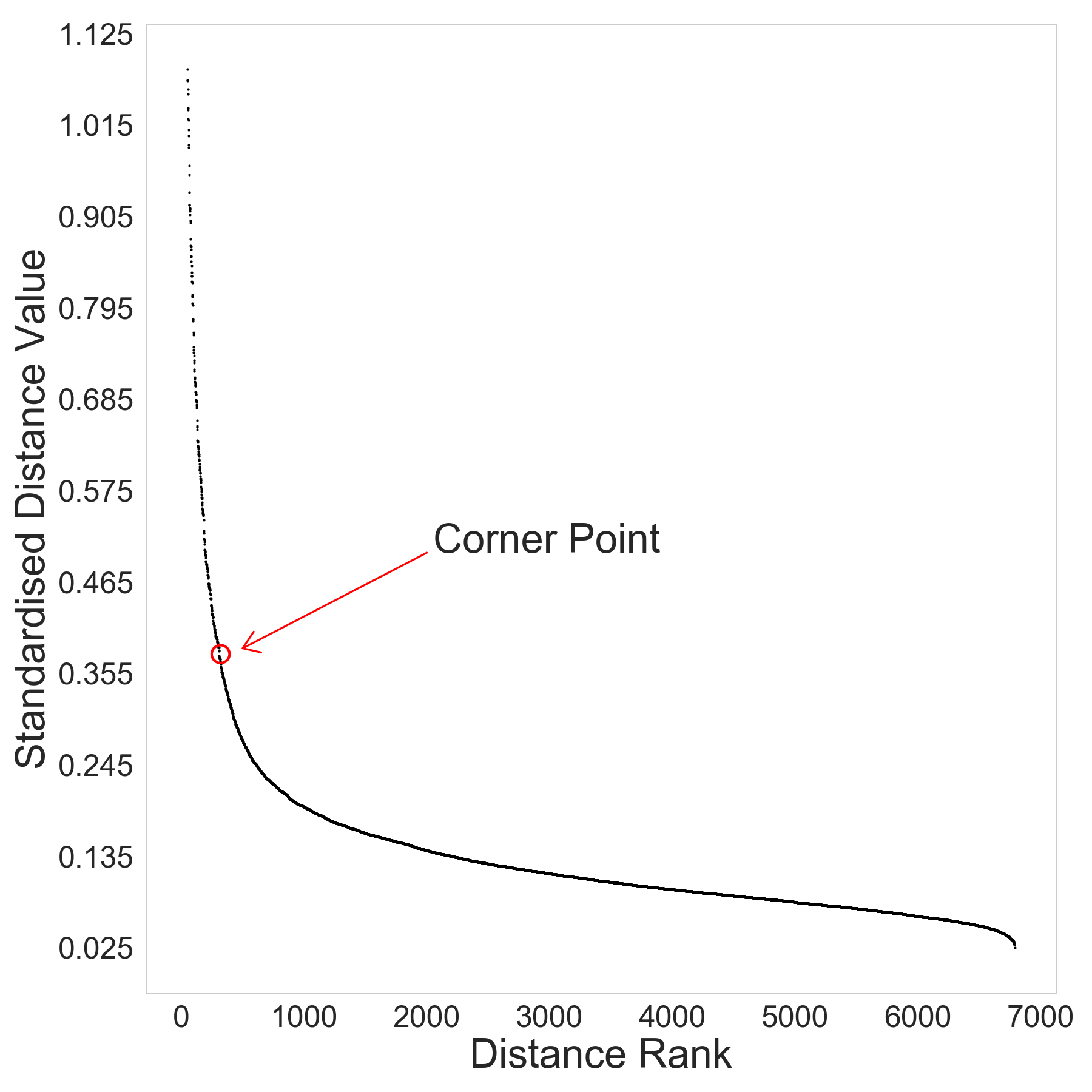}
    \caption{This is an example of a potential four-distance plot. In this figure, the y-axis represents the standardised distance values for the integrals of motion for each star. The x-axis shows the rank of each star with those
    with the largest standard distance ranked first. The arrow in the diagram points to the characteristic point where the plot becomes smooth, and this would be where we would extract the $\epsilon$ value. The data from this example comes from one of the Monte-Carlo iterations of the local stellar halo subset.}
    \label{fig: 4 dist example}
\end{figure}

Deriving the logic behind a 4-distance plot comes from the inherent characteristics of a clustered data set. In a clustered data set, the spacing between data values falls into two categories: the distances between members inside a cluster, and distances between data values which do not belong to the same cluster. Distances between two points inside a cluster are small relative to points outside the cluster. Additionally, points which are not part of the same cluster can also experience a significant variance in their distance from one another. Therefore, two populations of distances exist based on the different variances of the distances for a given data set. Determining the $\epsilon$ value is achieved through creating a ranked scatter plot of the distances between the points in the data set, with the highest rank given to the largest distance. The two populations can then be separated visually by finding the point in the plot where the scatter plot effectively becomes smooth. This junction represents the point of separation between the two populations, and we take this to be the value of $\epsilon$. Figure \ref{fig: 4 dist example} is an example of such a plot created during the process of determining the $\epsilon$ value for our stream search. The point at which the line changes from discrete to continuous is highlighted and taken as the $\epsilon$ value.

In order to take uncertainties into account, we create 4-distance plots for a random selection of 50 of the Monte-Carlo samples used in the search. We extract the $\epsilon$ from each of these 4-distance plots and take the average value. The average $\epsilon$ is a better measure of the clustering in the data set. Our choice to use 50 4-distance plots produced an accurate measure of $\epsilon$ without excessive time spent on calculations. The standardised $\epsilon$ values which we obtained for the replica data set were $\epsilon_{replica} = 0.286$, and for the restricted data $\epsilon_{restricted} = 0.306$. To provide context the ranges of the integrals of motion in this standardised form are provided in Table \ref{tab:Standardise Table}.

\begin{table}
    \centering
    \begin{tabular}{c|c|c|c}
        \hline
        \hline
         &  $L_z$ &  $L_{\perp}$ &  $Energy$ \\ \hline
        Max & 10.3 & 7.05 & 19.03\\ 
        Min & -5.48 & -1.32 & -1.15 \\
        \hline
    \end{tabular}
    \caption{The ranges of the integrals of motion when they have become standardised.}
    \label{tab:Standardise Table}
\end{table}

\subsection{Statistical Significance}
\label{sec:Significance}
We determine statistical significance for our DBSCAN clusters in integral of motion space through the use of the Milky Way observation simulator \textsc{galaxia} \citep{Sharma_Galaxia}. \textsc{galaxia} is capable of producing realistic synthetic representations of Galactic observations, enabling us to apply the same selections that we applied to \textit{Gaia} data to create a manufactured halo subset. With the simulated data, we can compare the clusters found by our method to the expected stellar density provided by \textsc{galaxia}'s output. If the number of stars in the space the cluster occupies in \textit{Gaia} is significantly higher than the simulated equivalent, we have confidence that the stream is a real kinematic structure.

We created three synthetic Galactic surveys and combined them in order to produce a data set with the same number density as \textit{Gaia}. Each synthetic survey was created by running \textsc{galaxia} with a different random seed integer, to create three plausible Galactic surveys. All default simulation parameters were used except for an apparent magnitude range of $M=4-13$ to mimic \textit{Gaia} data \citep{Gaia_B}. With the simulations complete, local halo subsets were created using the same method as described in Section \ref{sec:data}. Stars were then randomly drawn from this overabundant set to produce the same number of stars as the \textit{Gaia} halo. 

The statistical significance of the detected clusters was determined through the use of a multidimensional histogram, used to break the data into 3-dimensional cells of $L_{Z}$, $L_{\perp}$, and $E$. Bin sizes for the histogram were $\Delta L_{Z} = 400~km~kpc~s^{-1}$, $\Delta L_{\perp} = 200~km~kpc~s^{-1}$, and $\Delta E = 8500~km^{2}~s^{-2}$. Cells that did not have stream stars within them were then removed from this grid, leaving a 3-dimensional object that captured the volume in integral of motion space where the streams resided. Our choice of cell size split the integrals of motion into a $20\times20\times20$ grid of equally spaced cells. This cell size was small enough to be effective in capturing the general shape of the stream in integral of motion space, while not being so small as to prevent calculating useful Poisson statistics for the stars in each cell. A corresponding object comprised of the same cells was then created in the synthetic set, and the number of stars in those cells were counted. The numbers of stars contained within each object were defined as $N_i^{Gaia}$ and $N_i^{Model}$ respectively, while we defined the standard deviation of the synthetic cells to be 

\begin{equation}
    \sigma_i = \sqrt{N^{Model}_i}
\end{equation}

Our criterion for statistical significance is the same as is used in \citet{Williams_Galaxia}

\begin{equation}
    N^{Gaia}_i - N^{Model}_i > 4\sigma_i
\end{equation}

Upon passing this test, we declare the cluster in question to be a statistically significant grouping.

\section{Results}
\label{sec:results}
As described in section \ref{sec:data}, we are considering two different subsets of \textit{Gaia} data: the \say{replica} data set, which uses the same selection as in \citet{Koppel}, and the \say{restricted} data set, which is the replica data set with a correction applied to the parallax zero-point and only considering stars with five or greater independent radial velocity measurements. We will discuss the results of our stream searches in these two data sets separately.

\subsection{The replica data set}
\label{sec:Comparison Cut Search}
In the replica data set, we initially searched for streams out to 1 kpc of the Sun before incrementally increasing the search range out to 4 kpc. Our goal was to locate any streams residing in the local solar neighbourhood, and then test how far they extended beyond this distance. The purpose of this was to compare our results with \cite{Koppel} who found five: the H99 stream, initially discovered by \citet{Helmi_b}, and four additional stream candidates. We refer to these four a streams as Koppelman-\textit{Gaia}-Stream-1 (KGS-1), KGS-2, KGS-3 and KGS-4.

\begin{figure*}
    \centering
    \includegraphics[scale=0.23]{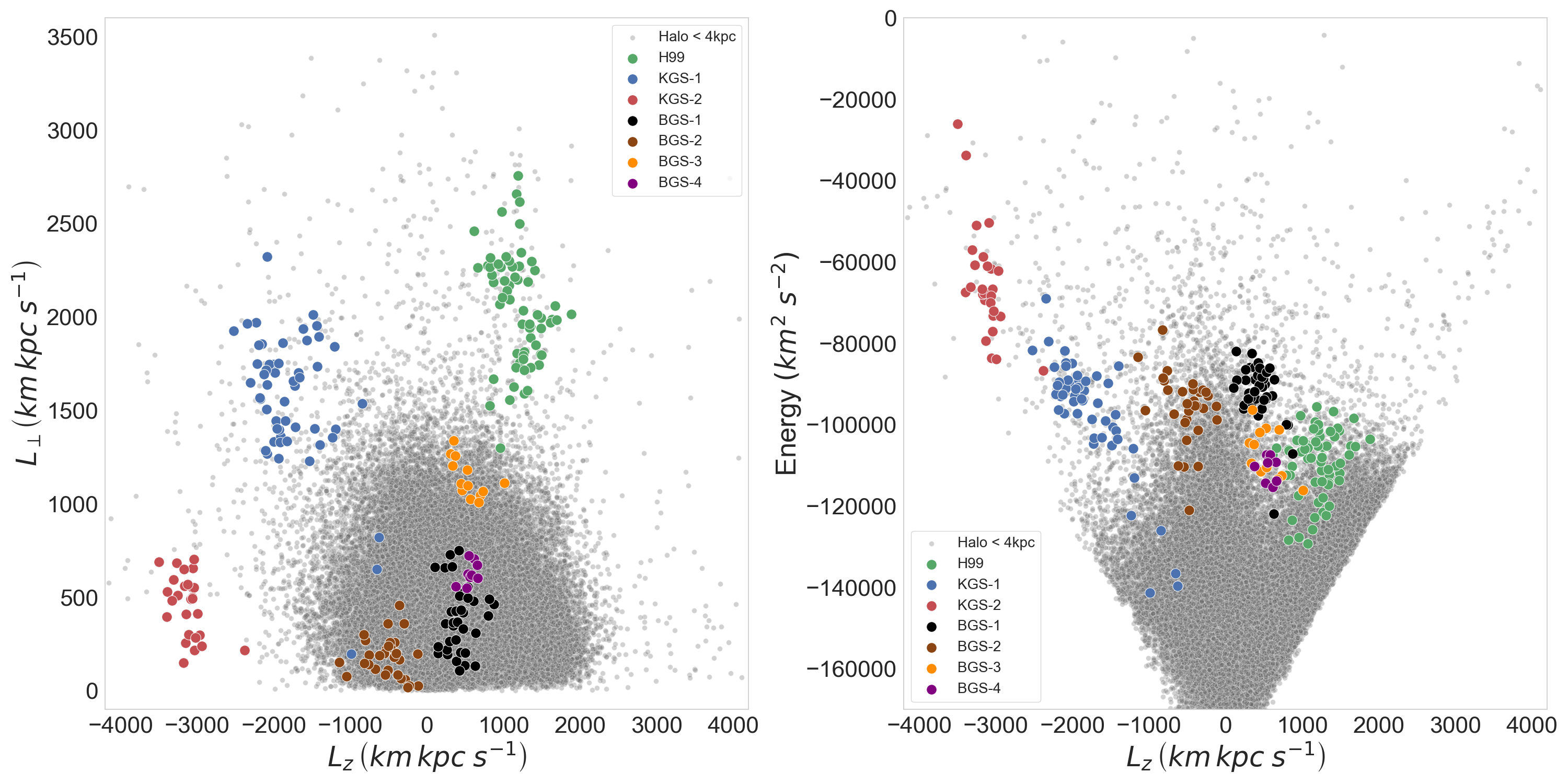}
    \caption{The integrals of motion distribution for the halo out to 4 kpc, for the \protect\say{replica} data set and the clusters found using our method. Coloured points represent clusters that the DBSCAN algorithm was able to resolve. The opaque grey data points in the background are the remaining distribution of halo stars which did not form any clusters.}
    \label{fig:Integrals_of_Motion_Clusters}
\end{figure*}

Through our efforts, we were able to resolve the H99, KGS-1, and KGS-2 streams using DBSCAN. Our method identified the majority of the previously reported stars said to belong to these streams and also indentifies new potential members. However, we were not able to resolve KGS-3 and KGS-4. Furthermore, it appears that some of the halo sub-structures found in \cite{Myeong_1,Myeong_2}, overlap with the H99 and KGS-1 streams. The sub-structures found in these publications used a cross match of \textit{Gaia} DR1 and the Sloan Digital Sky Survey data release 9, termed the SDSS-Gaia catalogue. In addition to confirming these known streams, we also identify four previously unknown sub-structures, which we will call Borsato-\textit{Gaia}-Stream-1 (BGS-1), BGS-2, BGS-3, and BGS-4. Table \ref{tab:Summary_table_1} lists the number of stars, distance range, means and standard deviations of the integrals of motion for each of the seven streams, and Figure \ref{fig:Integrals_of_Motion_Clusters} shows the integral of motion space for all stars in the replica data set, with the streams plotted in separate colours. Halo stars not marked as members of a stream by DBSCAN are plotted in the background as smaller opaque points, to better orient the reader.

\subsubsection{The Helmi stream}
The H99 stream remained statistically significant out to a distance of 1.5 kpc, with a total population of 64 stars. At distances greater than 1.5 kpc, our method could no longer resolve H99. The increased number of stars in the larger data set likely led to some overlap between the uncertainties of the stream stars and the surrounding population. This overlap would have prevented DBSCAN from creating a complete border. As a result, no cluster formed. We find that 55 of the 64 stars are common to the H99 stream found in \cite{Koppelman_2018_2}, though that study also found additional members as far away as 5 kpc. Of additional note is the fact that our method does not appear to cluster the H99 stars at higher energy values. The fixed $\epsilon$ value is likely preventing the inclusion of these stars as part of the cluster.

\begin{figure*}
    \centering
    \includegraphics[scale=0.23]{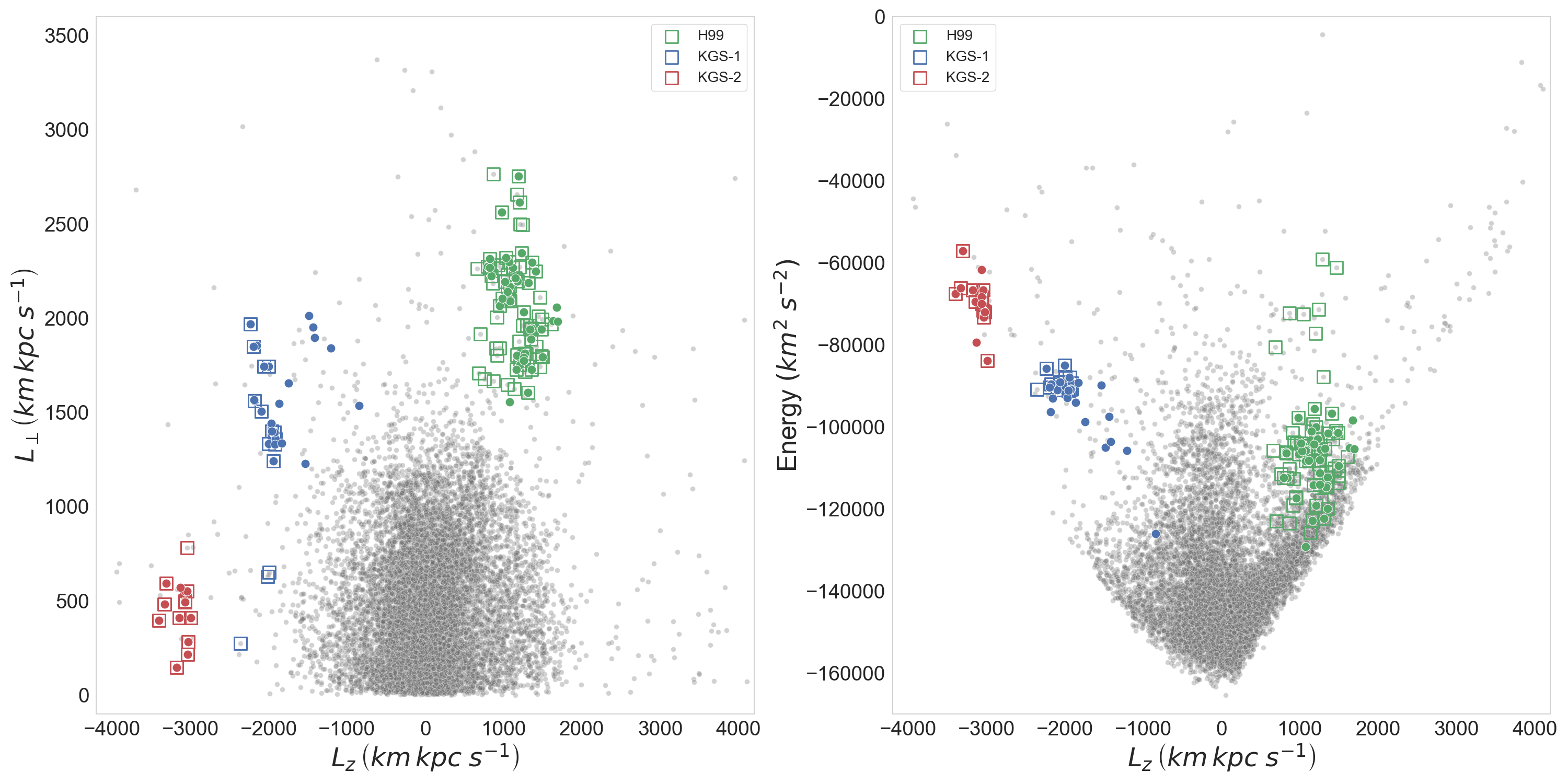}
    \caption{A comparison of our search results with the respective literature counterparts. Our stream members have been plotted as points on this figure, while the published stream members from \protect\cite{Koppel} have been plotted as open squares. A circle inside an individual box, therefore, represents stars identified in both works. The H99 stream stars in green represent all the stars we identify to a range of 1.5 kpc, which is as far as our search method was able to follow H99, and for the KG-streams we show all the stars in those streams out to a range of 1 kpc, the maximum range provided by \protect\cite{Koppel}.} 
    \label{fig:H99_Stars}
\end{figure*}

\subsubsection{The Koppelman-\textit{Gaia}-streams}
The two streams from \cite{Koppel} that we identify in the replica data set, KGS-1 and KGS-2, are depicted in blue and red in Figure \ref{fig:Integrals_of_Motion_Clusters}. In \citet{Koppel}, the search for streams was conducted within a 1 kpc range of the Sun, and so we restricted our initial search of the replica data set to the same range, and then repeated the search in larger volumes. For the KGS-1 stream, we identify 23 stars within 1 kpc of the Sun, of which 12 are also found in \citet{Koppel}. When extending the search range, we found KGS-1 extended out to 2 kpc from the Sun, and containing a total of 52 members. A noticeable peculiarity of this stream are the three stars that are offset from the rest of the group, mainly in $L_{\perp}$ and $E$. Follow-up analysis of these stars revealed that they have substantial error bars. The size of these errors was likely enough for DBSCAN to include these stars in some of the iterations, meaning that these stars are false positive detections.

For KGS-2 we identify 12 stream stars within 1 kpc of the Sun, of which 10 are also reported in \citet{Koppel}. Moreover, the KGS-2 stream remains coherent out to a range of 4 kpc with a final member count of 27. We suspect that the stream would remain statistically significant beyond this point but the increasing number of stars obtained by expanding the search range led to impractical calculation times for our stream searches. 

Figure \ref{fig:H99_Stars} shows the overlap of our stars (plotted as coloured circles) with the H99, KGS-1 and KGS-2 streams (shown as coloured squares). Stars included in both data sets have both a circle and a square. H99 contains all stream stars out to a range of 1.5 kpc, which was the maximum range our method could resolve the stars. KGS-1 and KGS-2 contain all stream stars out to 1 kpc, the maximum range that \cite{Koppel} reported. Restricting the ranges in this fashion allows one to fairly compare the results of each stream search, as the distances are consistent. 

\subsubsection{The Borsato-\textit{Gaia}-streams}
Similar to the H99 and KGS-1 and -2 streams, we first detected the BG-streams initially in the local solar neighbourhood, before examining how far they extended. The final results find that the BGS-1 stream extends out to 2 kpc, with a total of 38 members. BGS-2, BGS-3 and BGS-4 all stayed statistically significant out to 1.5 kpc and contained a final member count of 31, 13, and 9 members respectively.

\begin{table*}
    \centering
    \begin{tabular}{c|c|c|c|c|c|c|c}
        \hline
        \hline
        Stream &  H99 &  KGS-1 &  KGS-2 & BGS-1 & BGS-2 & BGS-3 & BGS-4\\ \hline
        Population & 64 & 52 & 27 & 38 & 31 & 13 & 9\\ 
        Range (kpc) & 1.5 & 2 & 4 & 2 & 1.5 & 1.5 & 1.5\\
        $L_{Z}$ \big($\frac{km~kpc}{s}$\big) & 1200 & -1800 & -3100 & 430 & -520 & 540 & 560\\
        $\sigma_{L_{Z}}$ & 250 & 420 & 200 & 180 & 240 & 190 & 80\\
        $L_{\perp}$ \big($\frac{km~kpc}{s}$\big) & 2000 & 1600 & 460 & 400 & 180 & 1100 & 630\\
        $\sigma_{L_{\perp}}$ & 300 & 350 & 160 & 180 & 100 & 100 & 60\\
        $E$ \big($\frac{km^2}{s^2}$\big) & -11000 & -97000 & -65000 & -92000 & -96000 & -110000 & -110000\\
        $\sigma_{E}$ & 8000 & 14000 & 13000 & 7000 & 8500 & 5700 & 2900\\\hline
    \end{tabular}
    \caption{Summary data of all streams we identify in the replica data set. Integrals of motion values are means.}
    \label{tab:Summary_table_1}
\end{table*}

\subsection{The restricted data set}
\label{sec:Quality Cut Search}
\begin{figure*}
    \centering
    \includegraphics[scale=0.23]{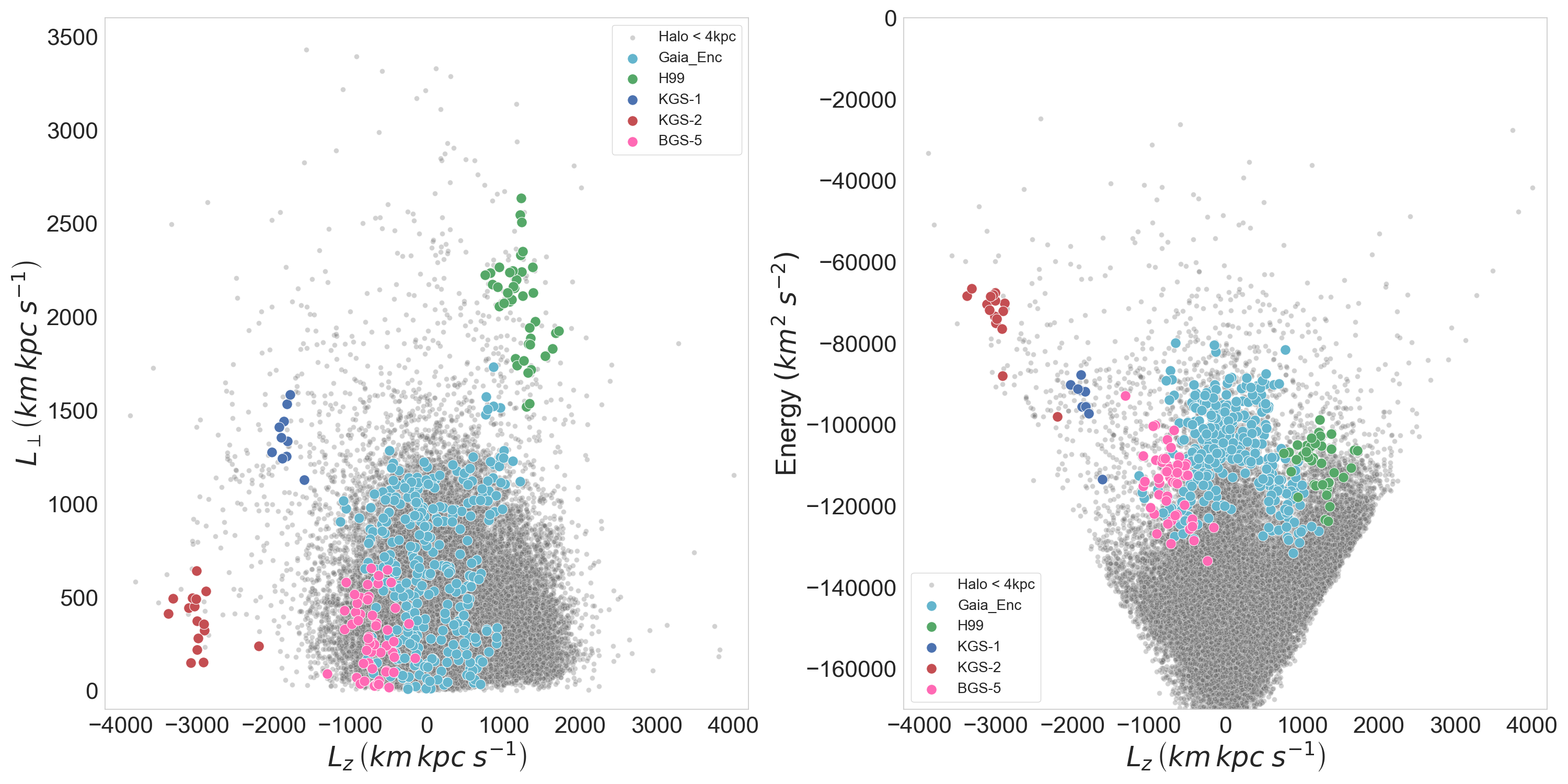}
    \caption{The integrals of motion distribution for the halo out to 4 kpc, with the quality cuts. Colour coding has remained consistent with Figure \protect\ref{fig:Integrals_of_Motion_Clusters}. Gaia-Enceladus is plotted in cyan and BGS-5 in pink. When comparing the before and after-effects of the new figures, one can see the loss in stellar members for H99, KGS-1, and KGS-2. BGS-5 and Gaia-Enceladus appear to overlap in some points, and it takes a plot in 3-dimensional integral of motion space to separate these two groups.}
    \label{fig:Integrals_of_motion_quality_cuts}
\end{figure*}

We applied the additional quality cuts, correcting for the global parallax offset and only accepting stars with $\geq 5$ radial velocity transits, and repeated the stream search method. Figure \ref{fig:Integrals_of_motion_quality_cuts} presents the results of our search attempts with this new data set. Similarly to the previous search, our method was able to locate the H99, KGS-1, and KGS-2 streams again. The original Borsato-\textit{Gaia}-Streams: BGS-1, -2, -3 and -4, were not detected in this new data set. The majority of the stars in these streams ($\sim 70\%$) were removed from the data set when the additional quality cut was made. As these streams were located in a much more dense region of integral of motion space, it is likely that the remaining stars did not form a dense enough cluster for the data-miner to relocate the streams in the restricted data set. 

However, we do identify two additional substructures located in the centre of the integral of motion plot that were not apparent in the replica data set. The larger of the two appears to be the Gaia-Enceladus dwarf galaxy remnant \citep{Helmi_Blob} as the cluster is located in the same region, and has its characteristic shape, a central bump in the $L_{Z}$ vs $E$ plot. The other cluster we have named BGS-5 because it does not correspond to any of the streams identified by \citet{Koppel}, or the Borsato-Gaia-Streams found in the replica data set. BGS-5 does appear to almost overlap with BGS-2, however, the two artefacts are distinctly separate when considering all three dimensions of the integrals of motion, and share no stars in common. 

The substructure we suspect is Gaia-Enceladus contains 319 members in total and remains statistically significant out to 1.5 kpc, while BGS-5 contains 53 members and remains significant out to 2 kpc. Table \ref{tab:Summary_table_2} presents these new streams with their average integrals of motion and standard deviations, along with the recovered H99, KGS-1 and -2 streams and their revised member totals. The three rogue stars originally included in KGS-1 were not included in the revised stream, implying the quality cut acted to remove these stars.

\begin{table*}
    \centering
    \begin{tabular}{c|c|c|c|c|c|c|c}
        \hline
        \hline
        Stream &  H99 &  KGS-1 &  KGS-2 & BGS-5 & Gaia-Enc\\ \hline
        Population & 41 & 10 & 16 & 53 & 319\\ 
        Range (kpc) & 1.5 & 2 & 4 & 2 & 1.5\\
        $L_{Z}$ \big($\frac{km~kpc}{s}$\big) & 1200 & -1900 & -3000 & -700 & 50\\
        $\sigma_{L_{Z}}$ & 480 & 210 & 100 & 240 & 210\\
        $L_{\perp}$ \big($\frac{km~kpc}{s}$\big) & 2100 & 1400 & 380 & 290 & 660\\
        $\sigma_{L_{\perp}}$ & 390 & 250 & 130 & 140 & 190\\
        $E$ \big($\frac{km^2}{s^2}$\big) & -11000 & -95000 & -74000 & -115000 & -110000\\
        $\sigma_{E}$ & 11000 & 5900 & 6800 & 8000 & 8200\\
        \hline
    \end{tabular}
    \caption{Summary data of all streams we identify in the second, more restricted, data set. Integrals of motion values are means.}
    \label{tab:Summary_table_2}
\end{table*}

\begin{figure*}
    \centering
    \includegraphics[scale=0.2]{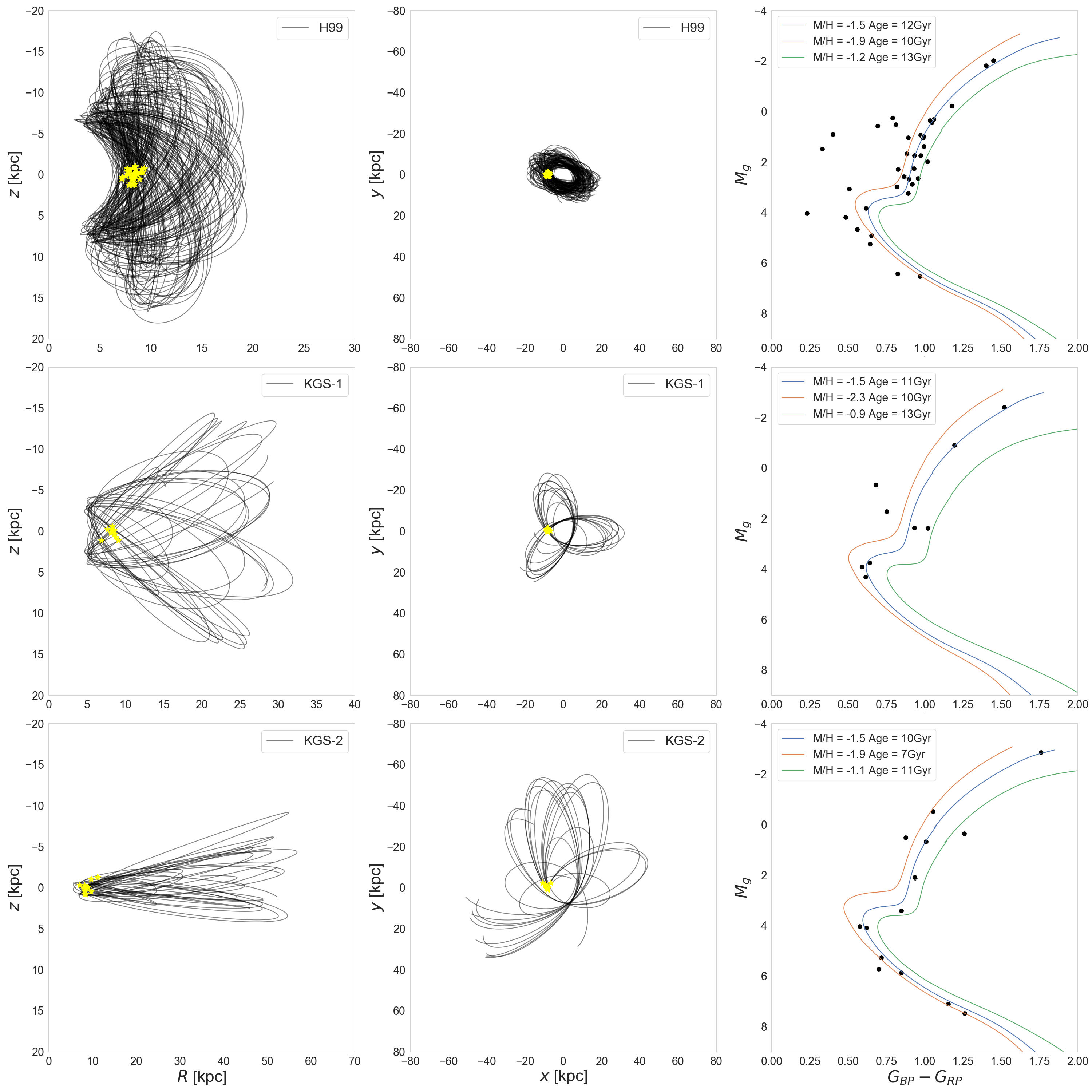}
    \caption{Orbits and photometry for the H99, KGS-1, and KGS-2 streams. The first column of plots shows the orbits in $R$ vs $Z$ projection, the second column shows the orbits in the $X$ vs $Y$ plane, and the third column is the absolute colour-magnitude diagram of each stream overlaid with Dartmouth isochrones \protect\citep{Isochrone_Model}. The orbit plots for the three streams are drawn on the same spatial scale for easy visual comparison. The yellow stars in the orbit plots represent the current positions of the stars as reported in the \textit{Gaia} DR2 catalogue. The blue isochrones represent the fitted isochrone path for this population of stars, with the green and orange isochrones denoting the bounds within the specified colour and magnitude ranges. The legend in each plot presents the ages and metallicities of the isochrones.}
    \label{fig:Re-Discovered_Streams}
\end{figure*}

\begin{figure*}
    \centering
    \includegraphics[scale=0.2]{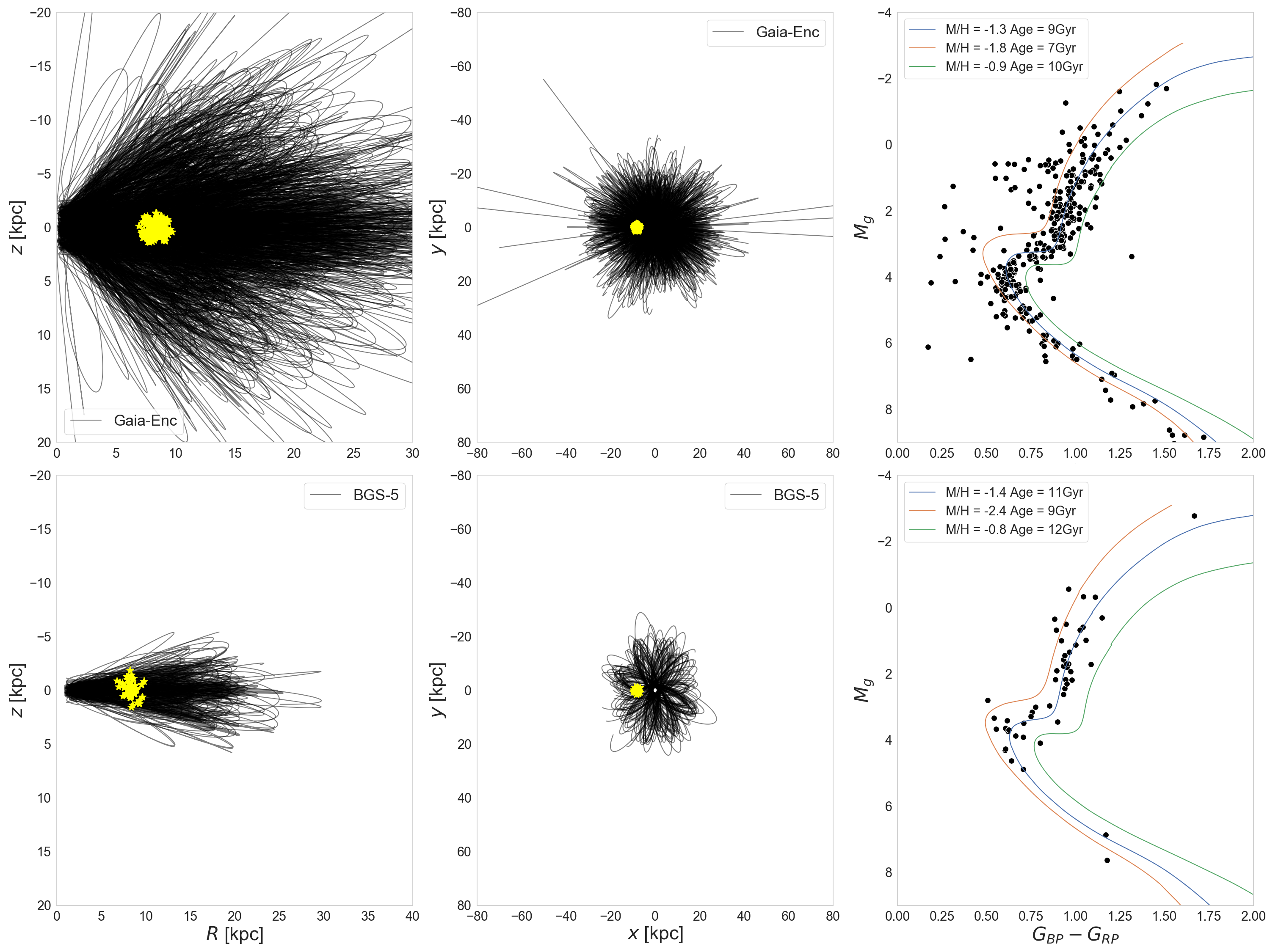}
    \caption{Orbits and photometry for the Gaia-Enceladus and BGS-5 streams, in the same style as Fig. \ref{fig:Re-Discovered_Streams}.}
    \label{fig:New_Streams}
\end{figure*}

Figures \ref{fig:Re-Discovered_Streams} and \ref{fig:New_Streams} show calculated orbits in Galactocentric $R$ vs $Z$ and $X$ vs $Y$ coordinates and absolute colour-magnitude diagrams for the stream members for the restricted data set. Equivalent figures for the replica data set can be found in Appendix \ref{app:replica-cmds}. Using each star's phase space coordinates as its initial conditions, the MilkyWayPotential class in \textsc{gala}, and a time step of 1 Myr, we integrate forward over each star's orbit for 1000 steps, which totals 1 Gyr. The result of this integration produces an Orbit object, which contains all the relevant parameters of the star's orbit at each time step \citep{Bovy_2016}. The orbit plots for each stream show each stream member's orbit for 1 Gyr, with the current positions of the stars shown in yellow.

Our colour-magnitude diagrams were made using the $G$, $G_{\rm BP}$ and $G_{\rm RP}$ magnitudes available from \textit{Gaia}. All stream stars showed appropriate \textit{Gaia} data quality, satisfying \texttt{phot\_bp\_rp\_excess\_factor} $< 1.3 + 0.06(G_{\rm BP}-G_{\rm RP})^2$ \citep{CMD_PAPER_GAIA}, with the exception of one star in the H99 stream, which had an \texttt{phot\_bp\_rp\_excess\_factor} $ = 2.3334208$ and has been removed. The \texttt{phot\_bp\_rp\_excess\_factor} delimits photometry that can be considered to be well-behaved in the \textit{Gaia} data set. We corrected for reddening and extinction using the publicly available extinction map from the Infrared Science Archive (IRSA) which uses reddening estimates provided by \cite{Reddening}, and converted the apparent magnitudes to absolute magnitudes using the \textit{Gaia} parallaxes \citep{CMD_PAPER_GAIA}. Dartmouth isochrones \citep{Isochrone_Model} were fitted by creating plausible tracks that ran through the body of the stellar population. Upper and lower bounds for the isochrones were created by varying the ages and metallicities of the isochrone tracks such that the average distance between the fitted isochrone and the isochrone bounds was no larger than $\Delta |M_v| = 0.35$ and $\Delta |G_{\rm BP}-G_{\rm RP}| = 0.1$. Any points that lay outside of the isochrone bounds were considered points of contamination. To quantify this contamination we took the ratio of stars that lay outside the isochrone bounds and divided by the total population of stars.

\begin{table}
    \centering
    \begin{tabular}{ll}
    \hline
    \hline
    \textit{Gaia} source ID & Stream \\
    \hline
    29380806120144384 & Enc \\
    30130265029073536 & Enc \\
    52624073910834176 & Enc \\
    53152251811120128 & Enc \\
    133539707282858880 & Enc \\
    \hline

    \end{tabular}
    \caption{\textit{Gaia} source ID numbers for our detected stream stars. The full machine-readable table is available online; these first few rows are shown here as a guide to form and content. \protect\say{Enc} is the label for Gaia-Enceladus.}
    \label{Gaia_5}
\end{table}

We begin by analysing the H99 stream orbits and colour-magnitude diagrams. H99 appears to have lost all of its spatial coherence and is dispersed throughout the Galaxy. This stream is considered to belong to a dwarf galaxy progenitor with an approximate mass of $\sim M^{8}_{\odot}$ and is projected to contribute as much as $\sim 10-14\%$ of the total stars in the Galactic halo in multiple stream-like structures \citep{Koppelman_2018_2}. Such multiple structures are likely the cause of the spatial dispersion in the orbits, as each stream-fragment may be on slightly different paths leading to a tangle of orbital paths that span a large portion of the sky. The colour-magnitude fit for this stream suggests an age of 12 Gyr and a metallicity of $[M/H] = -1.5$, which is in line with the findings of \cite{Koppelman_2018_2}. There are, however, stars which do not fall within the isochrone bounds. Using our dispersion approximation, we calculate that 32.5\% fall outside the isochrone bounds. As the population of this stream does not form a tight isochrone track, this progenitor is likely a dwarf galaxy containing a range of stellar populations.

KGS-1 has a significant vertical component to its orbit, with the extreme members of the stream reaching almost 20 kpc both above and below the Galactic plane while tracking a well defined elliptical path in the $X$ vs $Y$ plane. Analysing the colour-magnitude plots suggests that this stream it is old and metal-poor, with an approximate age of 11 Gyr and a metallicity of $[M/H] = -1.5$. The majority of the stars lie within the isochrone bounds with only 30\% of the stars lying outside.

The KGS-2 stream is also distinct from the other stream populations. The stars on the orbits in this stream fly out into the outer halo regions of our Galaxy before swinging back towards the Galactic centre. This stream satisfies a 10 Gyr fit with a metallicity of $[M/H] = -1.5$. 27\% of the stars in this colour-magnitude plot do not belong to this population of stars.

The next halo sub-structure that we will analyse is what we believe to be a remnant of the Gaia-Enceladus dwarf galaxy remnant. This sub-structure contained by far the most stars out of all the stream candidates found. The orbital plots in Figure \ref{fig:New_Streams} show that these stars will span the entire Galaxy, with a large fraction of them on highly radial orbits. The colour-magnitude diagram illustrates a clear population of stars with an approximate stellar age and metallicity of 9 Gyr and $[M/H] = -1.3$. The isochrone characteristics imply a slightly younger age than \cite{Helmi_Blob}; however, our metallicity measure is in agreement. Our dispersion measurements suggest that 24\% of the stars found in this stream do not match the isochrone fits. The colour-magnitude diagram clearly shows a coherent population of old, metal-poor stars with a characteristic spread that suggests that it is a remnant dwarf galaxy.

We consider the other new substructure we resolved in the restricted data set, the BGS-5 stream. All the stars in this stream appear to have highly radial orbits, manifested in the low $L_{Z}$ values of the stream. BGS-5 is also predicted to spend a significant amount of time orbiting through the thick disk of the Galaxy, seen in the $Z$ vs $R$ plot. Using our isochrone fitting procedure, we estimated the age of this stream of stars to be 11 Gyr with a metallicity value of $[M/H] = -1.4$. These values differ significantly with the Gaia-Enceladus fit, suggesting that they are two distinct populations of stars, even though they border each other quite closely in integral of motion space (see Figure \ref{fig:Integrals_of_motion_quality_cuts}). The dispersion is also quite low, with only 12\% of the stars falling outside the isochrone fits.

\begin{figure*}
    \centering
    \includegraphics[scale=0.23]{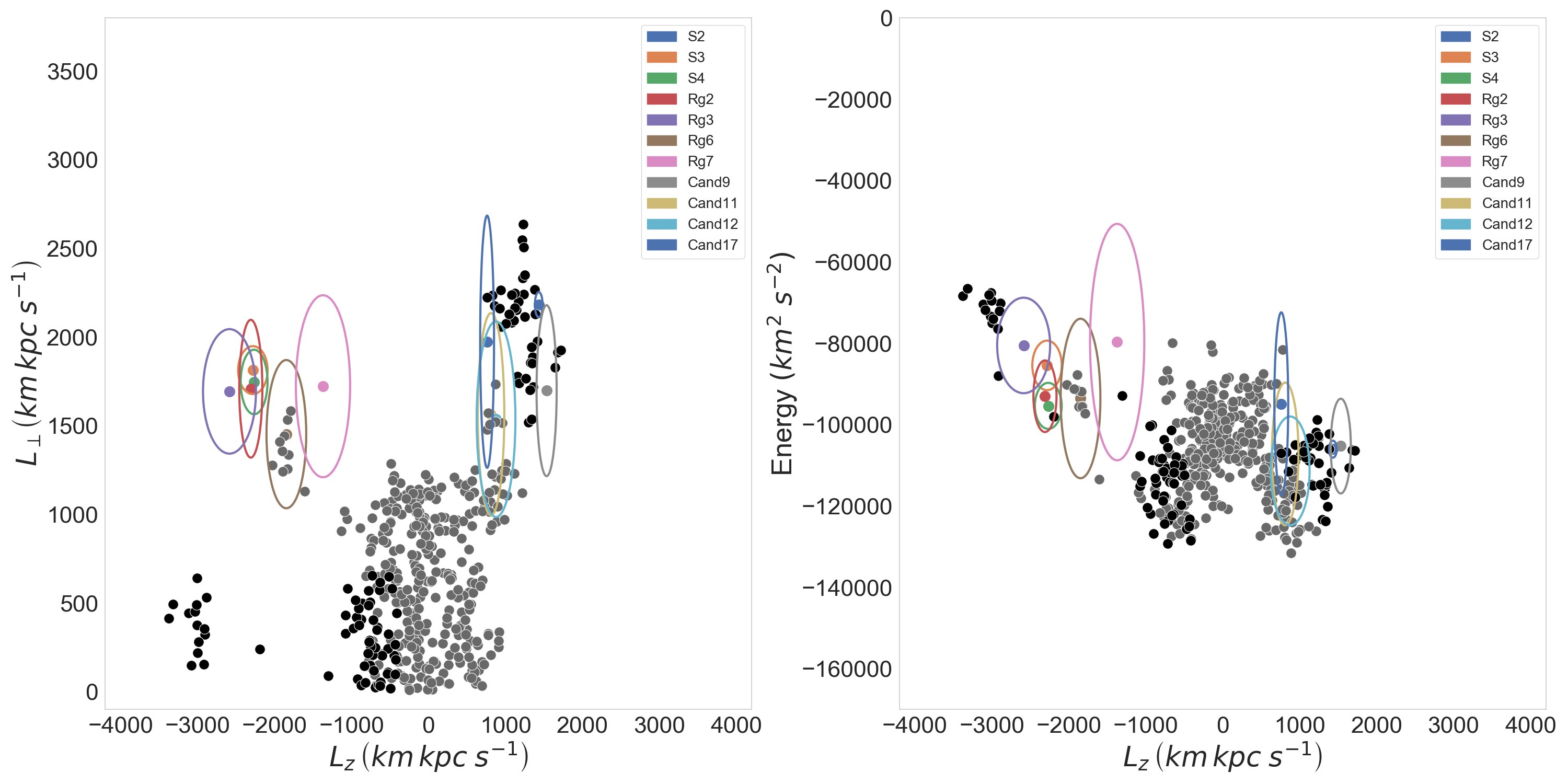}
    \caption{The halo sub-structures that potentially overlap with our stream candidates from \protect\cite{Myeong_1,Myeong_2}, using their stream names. There is a clear clustering of sub-structure close to the H99 and KGS-1 streams, which may indicate that these structures are all pieces of a larger progenitor system. Significantly, the Rg6 stream overlaps almost perfectly with the KGS-1 stream after the quality cut, implying that these stars form a single stream together.}
    \label{fig:Bell_Good}
\end{figure*}

\begin{figure*}
    \centering
    \includegraphics[scale=0.23]{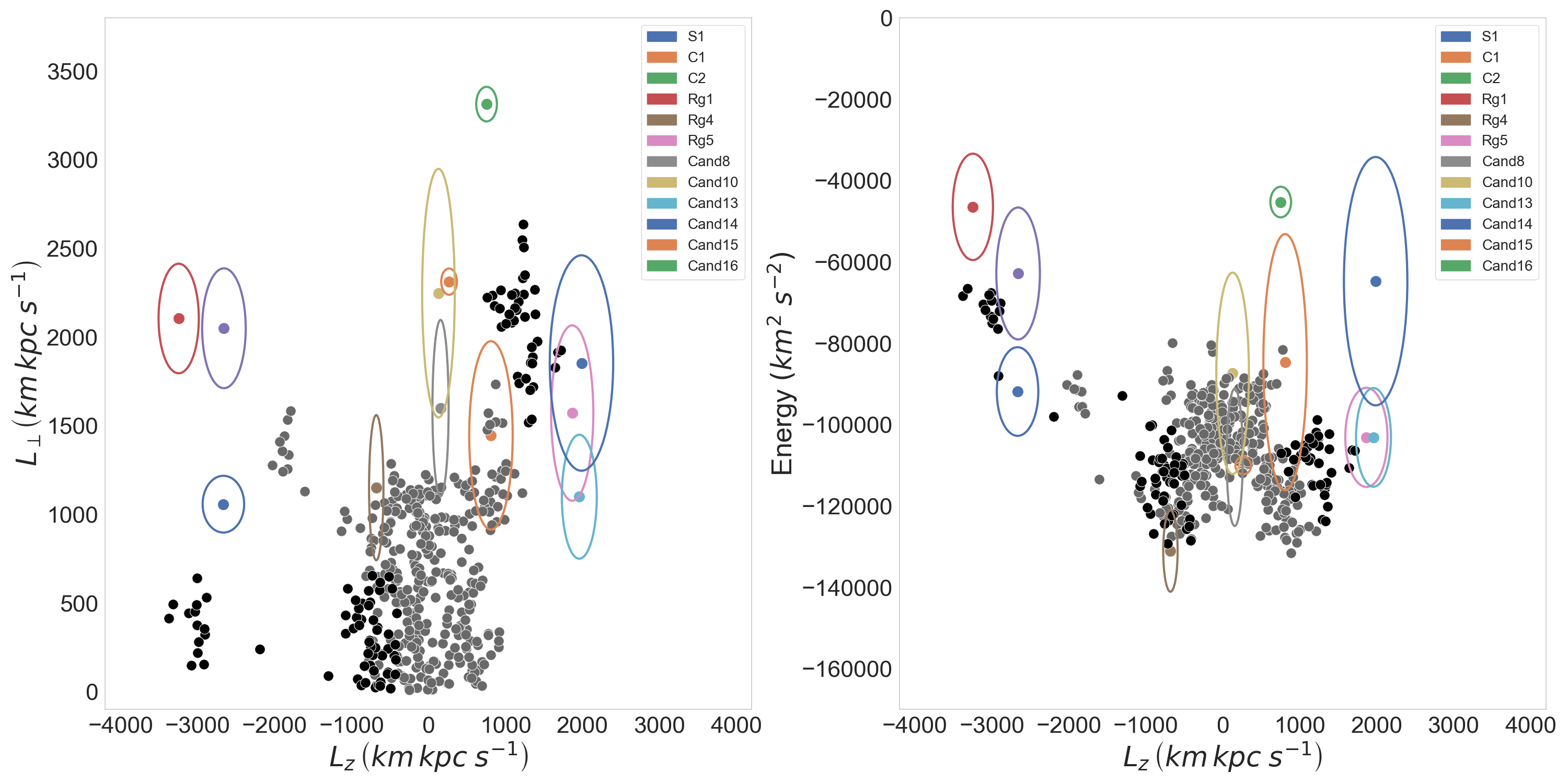}
    \caption{The halo sub-structures that do not overlap with our stream candidates from \protect\cite{Myeong_1,Myeong_2}, again using their stream names}.
    \label{fig:Bell_Bad}
\end{figure*}

Finally, we compare the streams that we have found in this paper with the halo sub-structures found in \cite{Myeong_1} and \cite{Myeong_2}, with the latter using the conserved actions of a star's orbit to find stream candidates. Figures \ref{fig:Bell_Good} and \ref{fig:Bell_Bad} indicate the locations of these structures with their integrals of motion, with the former plot illustrating likely sub-structures which share the same integrals of motion spaces as our stream. The integrals of motion for these streams were calculated using the phase space coordinates provided in \cite{Myeong_1} and \cite{Myeong_2}. The points in these plots represent the average $L_{Z}$, $L_{\perp}$ and $E$ values of these streams, while the ellipses show the standard deviation in those values, calculated from the errors provided in \cite{Myeong_1, Myeong_2}. All the streams we detected in this study have been plotted in grayscale, and all non-stream stars have been removed to better illustrate the positions of the Myeong streams. In Figure \ref{fig:Bell_Good}, a number of the streams found in \cite{Myeong_1} and \cite{Myeong_2} overlap with H99 and KGS-2. Of significant note is the Rg6 stream (where Rg stands for retrograde), which appears to almost entirely overlap with KGS-2 stream in the restricted data set, implying that there is a genuine chance that this sub-structure appears in both the \textit{Gaia} data set and the SDSS-\textit{Gaia} catalogue. It could very well be the case that Rg6 and KGS-2 are the same strucutre. The additional streams that lie close to H99 and KGS-2 without overlapping are likely to be fragments of a more extensive progenitor system. Dwarf galaxies are predicted to fragment during the accretion process \citep{Koppelman_2018_2}, rather than forming one monolithic cluster in integral of motion space. While only three of the streams identified in the replica data set were found after the transition to the restricted data set, the persistence of these features, and the consistency with some of the substructures identified by \citet{Myeong_1, Myeong_2}, give us confidence in the effectiveness and reliability of our method.

\section{Discussion}
\label{sec:discussion}
We have successfully demonstrated that searching for clusters of stars in the integral of motion space using data-mining techniques is a viable way to search for streams. Our method has a well-defined approach in locating and classifying streams, making the results easily reproducible. We now discuss the significance of these detections, as well as the issues and improvements we could apply in order to enhance our results. 

Of primary significance is the fact that our results using the restricted data set not only verified the existence of previously discovered streams but also have provided evidence that they extend past their ranges specified initially. The extension of the search distance outside the Solar neighbourhood has led us to understand that the H99 and KG-streams likely span a large portion of the Galaxy, which is in alignment with expectations \citep{Koppelman_2018_2}. It also may be the case that the additional stream members that we found in the \say{replica} data set but not the \say{restricted} set are also stream members. We plan to verify if this is true when the \textit{Gaia} DR3\footnote{Improved photometry and astrometry are planned to be released in late 2020. See \href{https://www.cosmos.esa.int/web/gaia/release}{ https://www.cosmos.esa.int/web/gaia/release} for details.} catalogue is released, and the number of RV transits for these stars has increased to satisfy the quality cut criteria. 

The quality cuts are responsible for our inability to resolve the original Borsato-Gaia-Streams in the restricted data set. We analysed the populations of the Gaia-Enceladus and BGS-5 streams and found that none of the stars in these clusters were members of the original BG-streams. The removal of a substantial number of the member stars after the quality cut was applied left a cluster with a density too low for the data miner to resolve. Conversely, Gaia-Enceladus and BGS-5 were not resolved in our initial search, likely because measurement errors on a number of lower quality stars in the data set washed out any coherent shape that DBSCAN would have found. With only the quality data remaining, this was no longer the case, and these coherent bodies of stars were found.

The orbital plots of these streams depict populations with a high degree of dispersion throughout the Galaxy. While the orbit paths for these stars do overlap in some instances, it is highly unlikely their existence would be detected if it were not for locating the streams using the stars' integrals of motion. The colour-magnitude plots show reasonably well defined populations, indicating that the stream stars are physically related and are not simply a selection of field stars.

Furthermore, our results provide evidence to suggest that dwarf galaxies fragment when accreted by their host galaxy, leading sub-structures to form within the larger cluster of the integrals of motion. We see this with the \cite{Myeong_1} streams, as their smaller sizes overlap with the areas where our streams are known to exist. This finding may imply that these structures originated from the same progenitor, and are in effect pieces of a larger whole. Fragmentation of a stream has already been predicted to have occurred with the H99 stream \citep{Koppelman_2018_2}. These findings suggest that while it is possible to locate stars of a similar origin using conserved quantities, it may be that other clusters of stars from the same progenitor too exist within the Galaxy, with integrals of motion which differ slightly. Further analysis of the properties of these streams are required to fully determine how and when they entered the Milky Way.

While our stream search technique has produced notable positive results, it could be improved. Our method of stream detection consistently produces results that differ slightly compared to the literature. DBSCAN failed to find two streams found in \cite{Koppel} KGS-3, and KGS-4, as well as failing to include all the stars in the streams that it did find. It is likely that a systematic difference in our search procedure is driving these differences in the results. The difference may not necessarily be a negative, as our search method can find clusters in the centre of the integrals of motion distribution where the stellar density is much higher. However, there are caveats to using DBSCAN. Our method can produce false positives for cluster membership. However, this can be mitigated if higher quality data is used. Furthermore, choosing a $MinPts$ value of four and relying on a 4-distance plot to select an $\epsilon$ value works well only if clustering size remains consistent throughout the data set, which generally is not the case for streams. Ideally, a data mining method that can vary its clustering range and produce clusters of multiple sizes would be optimal. Such a method would likely be more successful in capturing the stars with higher energy values in the H99 stream in Figure \ref{fig:Integrals_of_Motion_Clusters}. In addition to this, it is apparent that searching for clusters in integrals of motion is not immune to contamination of stars that are not part of the stream. More analysis is required when finding potential stream candidates in this way in order to determine whether the stars found do form a population of stars with the same origin. Caveats aside, while the algorithm may lack the required complexity to resolve all streams, and has a tendency to introduce contaminants, it is robust enough to produce consistent clustering in the data.

DBSCAN can, however, still consistently detect clusters even with some variation in the $\epsilon$ parameter. Our optimisation procedure changed the $\epsilon$ value when switching from the replica to the restricted data set, yet we were still capable of resolving the H99, KGS-1 and KGS-2 streams. The fact that results remained consistent for these streams means that DBSCAN is a robust search method, even when some leeway is permitted to vary the search parameters. The results of this study have shown that it is possible to locate stream candidates using the DBSCAN data miner to search for clusters in the measured integrals of motion of a star. While there are issues associated with the data miner, it is useful in locating clustering in integrals of motion which warrant legitimate detection of streams. 

\section*{Acknowledgements}
This work has made use of data from the European Space Agency (ESA) mission Gaia (http://www.cosmos.esa.int/gaia), processed by the Gaia Data Processing and Analysis Consortium (DPAC, 
http://www.cosmos.esa. int/web/gaia/dpac/consortium). Funding for the DPAC has been provided by national institutions, in particular the institutions participating in the Gaia Multilateral Agreement. Parts of this research were conducted by the Australian Research Council Centre of Excellence for All Sky Astrophysics in 3 Dimensions (ASTRO 3D), through project number CE170100013. JDS and SLM acknowledge the support of the Australian Research Council through Discovery Project grant DP180101791. NWB and SLM acknowledge the support of the UNSW Scientia Fellowship program.

For data mining and analysis, the following software packages have been used: \textsc{astropy} \citep{Astropy}, \textsc{gala} \citep{Gala}, \textsc{scikit-learn} \citep{scikit-learn}, \textsc{numpy} \citep{numpy}, \textsc{matplotlib} \citep{Matplotlib}, \textsc{seaborn} \citep{Seaborn}, \textsc{pandas} \citep{pandas}, and \textsc{topcat} \citep{TOPCAT}. This research has made use of the VizieR catalogue access tool, CDS, Strasbourg, France (DOI 10.26093/cds/vizier). The original description of the VizieR service was published in A\&AS 143, 23.

This research has also made use of the NASA/ IPAC Infrared Science Archive, which is operated by the Jet Propulsion Laboratory, California Institute of Technology, under contract with the National Aeronautics and Space Administration.




\bibliographystyle{mnras}
\bibliography{bib} 

\begin{thebibliography}{}
\makeatletter
\relax
\def\mn@urlcharsother{\let\do\@makeother \do\$\do\&\do\#\do\^\do\_\do\%\do\~}
\def\mn@doi{\begingroup\mn@urlcharsother \@ifnextchar [ {\mn@doi@}
  {\mn@doi@[]}}
\def\mn@doi@[#1]#2{\def\@tempa{#1}\ifx\@tempa\@empty \href
  {http://dx.doi.org/#2} {doi:#2}\else \href {http://dx.doi.org/#2} {#1}\fi
  \endgroup}
\def\mn@eprint#1#2{\mn@eprint@#1:#2::\@nil}
\def\mn@eprint@arXiv#1{\href {http://arxiv.org/abs/#1} {{\tt arXiv:#1}}}
\def\mn@eprint@dblp#1{\href {http://dblp.uni-trier.de/rec/bibtex/#1.xml}
  {dblp:#1}}
\def\mn@eprint@#1:#2:#3:#4\@nil{\def\@tempa {#1}\def\@tempb {#2}\def\@tempc
  {#3}\ifx \@tempc \@empty \let \@tempc \@tempb \let \@tempb \@tempa \fi \ifx
  \@tempb \@empty \def\@tempb {arXiv}\fi \@ifundefined
  {mn@eprint@\@tempb}{\@tempb:\@tempc}{\expandafter \expandafter \csname
  mn@eprint@\@tempb\endcsname \expandafter{\@tempc}}}

\bibitem[\protect\citeauthoryear{{Astropy Collaboration} et~al.,}{{Astropy
  Collaboration} et~al.}{2013}]{Astropy}
{Astropy Collaboration} et~al., 2013, \mn@doi [\aap]
  {10.1051/0004-6361/201322068}, \href
  {https://ui.adsabs.harvard.edu/\#abs/2013A&A...558A..33A} {558, A33}

\bibitem[\protect\citeauthoryear{{Bailer-Jones}}{{Bailer-Jones}}{2015}]{Parrallax-cut-off}
{Bailer-Jones} C. A.~L.,  2015, \mn@doi [Publications of the Astronomical
  Society of the Pacific] {10.1086/683116}, \href
  {https://ui.adsabs.harvard.edu/\#abs/2015PASP..127..994B} {127, 994}

\bibitem[\protect\citeauthoryear{{Balbinot} \& {Gieles}}{{Balbinot} \&
  {Gieles}}{2018}]{Balbinot_2018}
{Balbinot} E.,  {Gieles} M.,  2018, \mn@doi [\mnras] {10.1093/mnras/stx2708},
  \href {https://ui.adsabs.harvard.edu/\#abs/2018MNRAS.474.2479B} {474, 2479}

\bibitem[\protect\citeauthoryear{{Balbinot}, {Santiago}, {da Costa}, {Makler}
  \& {Maia}}{{Balbinot} et~al.}{2011}]{Balbinot_2017}
{Balbinot} E.,  {Santiago} B.~X.,  {da Costa} L.~N.,  {Makler} M.,   {Maia}
  M.~A.~G.,  2011, \mn@doi [\mnras] {10.1111/j.1365-2966.2011.19044.x}, \href
  {http://adsabs.harvard.edu/abs/2011MNRAS.416..393B} {416, 393}

\bibitem[\protect\citeauthoryear{{Belokurov} et~al.,}{{Belokurov}
  et~al.}{2006}]{Belokurov_2006}
{Belokurov} V.,  et~al., 2006, \mn@doi [\apjl] {10.1086/504797}, \href
  {http://adsabs.harvard.edu/abs/2006ApJ...642L.137B} {642, L137}

\bibitem[\protect\citeauthoryear{{Bernard} et~al.,}{{Bernard}
  et~al.}{2014}]{Bernard_2014}
{Bernard} E.~J.,  et~al., 2014, \mn@doi [\mnras] {10.1093/mnrasl/slu089}, \href
  {https://ui.adsabs.harvard.edu/\#abs/2014MNRAS.443L..84B} {443, L84}

\bibitem[\protect\citeauthoryear{Binney \& Tremaine}{Binney \&
  Tremaine}{2008}]{binney_tremaine_2008}
Binney J.,  Tremaine S.,  2008, Galactic dynamics.
Princeton University Press

\bibitem[\protect\citeauthoryear{{Blanton} et~al.,}{{Blanton}
  et~al.}{2017}]{SDSS}
{Blanton} M.~R.,  et~al., 2017, \mn@doi [\aj] {10.3847/1538-3881/aa7567}, \href
  {http://adsabs.harvard.edu/abs/2017AJ....154...28B} {154, 28}

\bibitem[\protect\citeauthoryear{{Bonaca}, {Conroy}, {Wetzel}, {Hopkins}  \&
  {Kere{\v s}}}{{Bonaca} et~al.}{2017}]{Bonaca_2017}
{Bonaca} A.,  {Conroy} C.,  {Wetzel} A.,  {Hopkins} P.~F.,   {Kere{\v s}} D.,
  2017, \mn@doi [\apj] {10.3847/1538-4357/aa7d0c}, \href
  {http://adsabs.harvard.edu/abs/2017ApJ...845..101B} {845, 101}

\bibitem[\protect\citeauthoryear{{Bose}, {Ginsburg}  \& {Loeb}}{{Bose}
  et~al.}{2018}]{Bose_2018}
{Bose} S.,  {Ginsburg} I.,   {Loeb} A.,  2018, \mn@doi [\apj]
  {10.3847/2041-8213/aac48c}, \href
  {https://ui.adsabs.harvard.edu/\#abs/2018ApJ...859L..13B} {859, L13}

\bibitem[\protect\citeauthoryear{{Bovy}}{{Bovy}}{2015}]{Bovy_2015}
{Bovy} J.,  2015, \mn@doi [The Astrophysical Journal Supplement Series]
  {10.1088/0067-0049/216/2/29}, \href
  {https://ui.adsabs.harvard.edu/\#abs/2015ApJS..216...29B} {216, 29}

\bibitem[\protect\citeauthoryear{{Bovy}, {Bahmanyar}, {Fritz}  \&
  {Kallivayalil}}{{Bovy} et~al.}{2016}]{Bovy_2016}
{Bovy} J.,  {Bahmanyar} A.,  {Fritz} T.~K.,   {Kallivayalil} N.,  2016, \mn@doi
  [\apj] {10.3847/1538-4357/833/1/31}, \href
  {https://ui.adsabs.harvard.edu/\#abs/2016ApJ...833...31B} {833, 31}

\bibitem[\protect\citeauthoryear{{Bowden}, {Belokurov}  \& {Evans}}{{Bowden}
  et~al.}{2015}]{Bowden_2015}
{Bowden} A.,  {Belokurov} V.,   {Evans} N.~W.,  2015, \mn@doi [\mnras]
  {10.1093/mnras/stv285}, \href
  {https://ui.adsabs.harvard.edu/abs/2015MNRAS.449.1391B} {449, 1391}

\bibitem[\protect\citeauthoryear{{Buder} et~al.,}{{Buder}
  et~al.}{2018}]{GALAH_DR2}
{Buder} S.,  et~al., 2018, \mn@doi [\mnras] {10.1093/mnras/sty1281}, \href
  {https://ui.adsabs.harvard.edu/abs/2018MNRAS.478.4513B} {478, 4513}

\bibitem[\protect\citeauthoryear{{Bullock} \& {Johnston}}{{Bullock} \&
  {Johnston}}{2005}]{Bullock_Johnston_2010}
{Bullock} J.~S.,  {Johnston} K.~V.,  2005, \mn@doi [\apj] {10.1086/497422},
  \href {https://ui.adsabs.harvard.edu/abs/2005ApJ...635..931B} {635, 931}

\bibitem[\protect\citeauthoryear{{Casey} et~al.,}{{Casey}
  et~al.}{2017}]{RAVEon}
{Casey} A.~R.,  et~al., 2017, \mn@doi [\apj] {10.3847/1538-4357/aa69c2}, \href
  {https://ui.adsabs.harvard.edu/abs/2017ApJ...840...59C} {840, 59}

\bibitem[\protect\citeauthoryear{{Chen} et~al.,}{{Chen}
  et~al.}{2001}]{Height_From_Galc_Cen}
{Chen} B.,  et~al., 2001, \mn@doi [\apj] {10.1086/320647}, \href
  {https://ui.adsabs.harvard.edu/abs/2001ApJ...553..184C} {553, 184}

\bibitem[\protect\citeauthoryear{{Chen}, {D'Onghia}, {Pardy}, {Pasquali},
  {Bertelli Motta}, {Hanlon}  \& {Grebel}}{{Chen} et~al.}{2018}]{Chen_2018}
{Chen} B.,  {D'Onghia} E.,  {Pardy} S.~A.,  {Pasquali} A.,  {Bertelli Motta}
  C.,  {Hanlon} B.,   {Grebel} E.~K.,  2018, \mn@doi [\apj]
  {10.3847/1538-4357/aac325}, \href
  {https://ui.adsabs.harvard.edu/\#abs/2018ApJ...860...70C} {860, 70}

\bibitem[\protect\citeauthoryear{{Davis}, {Efstathiou}, {Frenk}  \&
  {White}}{{Davis} et~al.}{1985}]{friend-of-friends}
{Davis} M.,  {Efstathiou} G.,  {Frenk} C.~S.,   {White} S.~D.~M.,  1985,
  \mn@doi [\apj] {10.1086/163168}, \href
  {https://ui.adsabs.harvard.edu/abs/1985ApJ...292..371D} {292, 371}

\bibitem[\protect\citeauthoryear{{Dotter}, {Chaboyer}, {Jevremovi{\'c}},
  {Kostov}, {Baron}  \& {Ferguson}}{{Dotter} et~al.}{2008}]{Isochrone_Model}
{Dotter} A.,  {Chaboyer} B.,  {Jevremovi{\'c}} D.,  {Kostov} V.,  {Baron} E.,
  {Ferguson} J.~W.,  2008, \mn@doi [\apjs] {10.1086/589654}, \href
  {https://ui.adsabs.harvard.edu/abs/2008ApJS..178...89D} {178, 89}

\bibitem[\protect\citeauthoryear{Ester, Kriegel, Sander  \& Xu}{Ester
  et~al.}{1996}]{Ester_1996}
Ester M.,  Kriegel H.-P.,  Sander J.,   Xu X.,  1996, Association for the
  Advancement of Artificial Intelligence, pp 226--231

\bibitem[\protect\citeauthoryear{{Gaia Collaboration} et~al.,}{{Gaia
  Collaboration} et~al.}{2018a}]{Gaia_B}
{Gaia Collaboration} et~al., 2018a, \mn@doi [\aap]
  {10.1051/0004-6361/201833051}, \href
  {https://ui.adsabs.harvard.edu/\#abs/2018A&A...616A...1G} {616, A1}

\bibitem[\protect\citeauthoryear{{Gaia Collaboration} et~al.,}{{Gaia
  Collaboration} et~al.}{2018b}]{CMD_PAPER_GAIA}
{Gaia Collaboration} et~al., 2018b, \mn@doi [\aap]
  {10.1051/0004-6361/201832843}, \href
  {https://ui.adsabs.harvard.edu/abs/2018A&A...616A..10G} {616, A10}

\bibitem[\protect\citeauthoryear{{Gillessen}, {Eisenhauer}, {Trippe}, {Alexand
  er}, {Genzel}, {Martins}  \& {Ott}}{{Gillessen}
  et~al.}{2009}]{Dist_From_Gal_Center}
{Gillessen} S.,  {Eisenhauer} F.,  {Trippe} S.,  {Alexand er} T.,  {Genzel} R.,
   {Martins} F.,   {Ott} T.,  2009, \mn@doi [\apj]
  {10.1088/0004-637X/692/2/1075}, \href
  {https://ui.adsabs.harvard.edu/abs/2009ApJ...692.1075G} {692, 1075}

\bibitem[\protect\citeauthoryear{{Grillmair} \& {Dionatos}}{{Grillmair} \&
  {Dionatos}}{2006}]{Grillmair_and_Dinotos_2006}
{Grillmair} C.~J.,  {Dionatos} O.,  2006, \mn@doi [\apj] {10.1086/505111},
  \href {https://ui.adsabs.harvard.edu/\#abs/2006ApJ...643L..17G} {643, L17}

\bibitem[\protect\citeauthoryear{{Grillmair}, {Freeman}, {Irwin}  \&
  {Quinn}}{{Grillmair} et~al.}{1995}]{Grillmair_Freeman_1995}
{Grillmair} C.~J.,  {Freeman} K.~C.,  {Irwin} M.,   {Quinn} P.~J.,  1995,
  \mn@doi [\aj] {10.1086/117470}, \href
  {https://ui.adsabs.harvard.edu/abs/1995AJ....109.2553G} {109, 2553}

\bibitem[\protect\citeauthoryear{{Helmi}, {Zhao}  \& {de Zeeuw}}{{Helmi}
  et~al.}{1999a}]{Helmi_b}
{Helmi} A.,  {Zhao} H.,   {de Zeeuw} T.,  1999a, in {Gibson} B.~K.,  {Axelrod}
  R.~S.,   {Putman} M.~E.,  eds,  Astronomical Society of the Pacific
  Conference Series Vol. 165, The Third Stromlo Symposium: The Galactic Halo.
  p.~125 (\mn@eprint {arXiv} {astro-ph/9811109})

\bibitem[\protect\citeauthoryear{{Helmi}, {White}, {de Zeeuw}  \&
  {Zhao}}{{Helmi} et~al.}{1999b}]{Helmi_a}
{Helmi} A.,  {White} S. D.~M.,  {de Zeeuw} P.~T.,   {Zhao} H.,  1999b, \mn@doi
  [\nat] {10.1038/46980}, \href
  {https://ui.adsabs.harvard.edu/\#abs/1999Natur.402...53H} {402, 53}

\bibitem[\protect\citeauthoryear{{Helmi}, {Veljanoski}, {Breddels}, {Tian}  \&
  {Sales}}{{Helmi} et~al.}{2017}]{Helmi_2017}
{Helmi} A.,  {Veljanoski} J.,  {Breddels} M.~A.,  {Tian} H.,   {Sales} L.~V.,
  2017, \mn@doi [\aap] {10.1051/0004-6361/201629990}, \href
  {https://ui.adsabs.harvard.edu/\#abs/2017A&A...598A..58H} {598, A58}

\bibitem[\protect\citeauthoryear{{Helmi}, {Babusiaux}, {Koppelman}, {Massari},
  {Veljanoski}  \& {Brown}}{{Helmi} et~al.}{2018}]{Helmi_Blob}
{Helmi} A.,  {Babusiaux} C.,  {Koppelman} H.~H.,  {Massari} D.,  {Veljanoski}
  J.,   {Brown} A. G.~A.,  2018, \mn@doi [\nat] {10.1038/s41586-018-0625-x},
  \href {https://ui.adsabs.harvard.edu/\#abs/2018Natur.563...85H} {563, 85}

\bibitem[\protect\citeauthoryear{Hernquist}{Hernquist}{1990}]{hernquist_1990}
Hernquist L.,  1990, \mn@doi [The Astrophysical Journal] {10.1086/168845}, 356,
  359

\bibitem[\protect\citeauthoryear{Hunter}{Hunter}{2007}]{Matplotlib}
Hunter J.~D.,  2007, Computing In Science \& Engineering, 9, 90

\bibitem[\protect\citeauthoryear{{Ibata}, {Irwin}, {Lewis}, {Ferguson}  \&
  {Tanvir}}{{Ibata} et~al.}{2001}]{Ibata_2001}
{Ibata} R.,  {Irwin} M.,  {Lewis} G.,  {Ferguson} A.~M.~N.,   {Tanvir} N.,
  2001, \nat, \href {http://adsabs.harvard.edu/abs/2001Natur.412...49I} {412,
  49}

\bibitem[\protect\citeauthoryear{{Ibata}, {Malhan}, {Martin}  \&
  {Starkenburg}}{{Ibata} et~al.}{2018}]{Malhan_4}
{Ibata} R.~A.,  {Malhan} K.,  {Martin} N.~F.,   {Starkenburg} E.,  2018,
  \mn@doi [\apj] {10.3847/1538-4357/aadba3}, \href
  {https://ui.adsabs.harvard.edu/\#abs/2018ApJ...865...85I} {865, 85}

\bibitem[\protect\citeauthoryear{{Ibata}, {Malhan}  \& {Martin}}{{Ibata}
  et~al.}{2019}]{Ibata_2019_Malhan_5}
{Ibata} R.~A.,  {Malhan} K.,   {Martin} N.~F.,  2019, \mn@doi [\apj]
  {10.3847/1538-4357/ab0080}, \href
  {https://ui.adsabs.harvard.edu/\#abs/2019ApJ...872..152I} {872, 152}

\bibitem[\protect\citeauthoryear{Johnson \& Soderblom}{Johnson \&
  Soderblom}{1987}]{johnson_soderblom_1987}
Johnson D. R.~H.,  Soderblom D.~R.,  1987, \mn@doi [The Astronomical Journal]
  {10.1086/114370}, 93, 864

\bibitem[\protect\citeauthoryear{{Johnston}, {Hernquist}  \&
  {Bolte}}{{Johnston} et~al.}{1996}]{Johnston_1996}
{Johnston} K.~V.,  {Hernquist} L.,   {Bolte} M.,  1996, \mn@doi [\apj]
  {10.1086/177418}, \href {http://adsabs.harvard.edu/abs/1996ApJ...465..278J}
  {465, 278}

\bibitem[\protect\citeauthoryear{{Klement}}{{Klement}}{2010}]{Klement_2010}
{Klement} R.~J.,  2010, \mn@doi [Astronomy and Astrophysics Review]
  {10.1007/s00159-010-0034-0}, \href
  {https://ui.adsabs.harvard.edu/\#abs/2010A&ARv..18..567K} {18, 567}

\bibitem[\protect\citeauthoryear{{Klement} et~al.,}{{Klement}
  et~al.}{2009}]{Klement_2009}
{Klement} R.,  et~al., 2009, \mn@doi [\apj] {10.1088/0004-637X/698/1/865},
  \href {https://ui.adsabs.harvard.edu/\#abs/2009ApJ...698..865K} {698, 865}

\bibitem[\protect\citeauthoryear{{Koposov}, {Rix}  \& {Hogg}}{{Koposov}
  et~al.}{2010}]{Koposov_2010}
{Koposov} S.~E.,  {Rix} H.-W.,   {Hogg} D.~W.,  2010, \mn@doi [\apj]
  {10.1088/0004-637X/712/1/260}, \href
  {https://ui.adsabs.harvard.edu/abs/2010ApJ...712..260K} {712, 260}

\bibitem[\protect\citeauthoryear{{Koppelman}, {Helmi}  \&
  {Veljanoski}}{{Koppelman} et~al.}{2018}]{Koppel}
{Koppelman} H.,  {Helmi} A.,   {Veljanoski} J.,  2018, \mn@doi [\apj]
  {10.3847/2041-8213/aac882}, \href
  {https://ui.adsabs.harvard.edu/\#abs/2018ApJ...860L..11K} {860, L11}

\bibitem[\protect\citeauthoryear{{Koppelman}, {Helmi}, {Massari}, {Roelenga}
  \& {Bastian}}{{Koppelman} et~al.}{2019}]{Koppelman_2018_2}
{Koppelman} H.~H.,  {Helmi} A.,  {Massari} D.,  {Roelenga} S.,   {Bastian} U.,
  2019, \mn@doi [\aap] {10.1051/0004-6361/201834769}, \href
  {https://ui.adsabs.harvard.edu/abs/2019A&A...625A...5K} {625, A5}

\bibitem[\protect\citeauthoryear{{Kunder} et~al.,}{{Kunder}
  et~al.}{2017}]{RAVE}
{Kunder} A.,  et~al., 2017, \mn@doi [\aj] {10.3847/1538-3881/153/2/75}, \href
  {https://ui.adsabs.harvard.edu/abs/2017AJ....153...75K} {153, 75}

\bibitem[\protect\citeauthoryear{{Law} \& {Majewski}}{{Law} \&
  {Majewski}}{2010}]{Law_and_Maj_2010}
{Law} D.~R.,  {Majewski} S.~R.,  2010, \mn@doi [\apj]
  {10.1088/0004-637X/714/1/229}, \href
  {https://ui.adsabs.harvard.edu/\#abs/2010ApJ...714..229L} {714, 229}

\bibitem[\protect\citeauthoryear{{Law}, {Johnston}  \& {Majewski}}{{Law}
  et~al.}{2005}]{Law_2005}
{Law} D.~R.,  {Johnston} K.~V.,   {Majewski} S.~R.,  2005, \mn@doi [\apj]
  {10.1086/426779}, \href {http://adsabs.harvard.edu/abs/2005ApJ...619..807L}
  {619, 807}

\bibitem[\protect\citeauthoryear{{Leung} \& {Bovy}}{{Leung} \&
  {Bovy}}{2019}]{Leung_Bovy_Parallax_Offset}
{Leung} H.~W.,  {Bovy} J.,  2019, \mn@doi [\mnras] {10.1093/mnras/stz2245},
  \href {https://ui.adsabs.harvard.edu/abs/2019MNRAS.tmp.2167L} {p.~2167}

\bibitem[\protect\citeauthoryear{Lloyd}{Lloyd}{1982}]{kmeans}
Lloyd S.~P.,  1982, IEEE Transactions on Information Theory, 28, 129

\bibitem[\protect\citeauthoryear{{Luo} et~al.,}{{Luo} et~al.}{2015}]{LAMOST}
{Luo} A.~L.,  et~al., 2015, arXiv e-prints, \href
  {https://ui.adsabs.harvard.edu/abs/2015arXiv150501570L} {p. arXiv:1505.01570}

\bibitem[\protect\citeauthoryear{{Malhan} \& {Ibata}}{{Malhan} \&
  {Ibata}}{2018}]{Malhan_1}
{Malhan} K.,  {Ibata} R.~A.,  2018, \mn@doi [\mnras] {10.1093/mnras/sty912},
  \href {https://ui.adsabs.harvard.edu/\#abs/2018MNRAS.477.4063M} {477, 4063}

\bibitem[\protect\citeauthoryear{{Malhan} \& {Ibata}}{{Malhan} \&
  {Ibata}}{2019}]{Malhan_Ibata_2019}
{Malhan} K.,  {Ibata} R.~A.,  2019, \mn@doi [\mnras] {10.1093/mnras/stz1035},
  \href {https://ui.adsabs.harvard.edu/abs/2019MNRAS.486.2995M} {486, 2995}

\bibitem[\protect\citeauthoryear{{Malhan}, {Ibata}, {Goldman}, {Martin},
  {Magnier}  \& {Chambers}}{{Malhan} et~al.}{2018a}]{Malhan_3}
{Malhan} K.,  {Ibata} R.~A.,  {Goldman} B.,  {Martin} N.~F.,  {Magnier} E.,
  {Chambers} K.,  2018a, \mn@doi [\mnras] {10.1093/mnras/sty1338}, \href
  {https://ui.adsabs.harvard.edu/\#abs/2018MNRAS.478.3862M} {478, 3862}

\bibitem[\protect\citeauthoryear{{Malhan}, {Ibata}  \& {Martin}}{{Malhan}
  et~al.}{2018b}]{Malhan_2}
{Malhan} K.,  {Ibata} R.~A.,   {Martin} N.~F.,  2018b, \mn@doi [\mnras]
  {10.1093/mnras/sty2474}, \href
  {https://ui.adsabs.harvard.edu/\#abs/2018MNRAS.481.3442M} {481, 3442}

\bibitem[\protect\citeauthoryear{{Marchetti}, {Rossi}  \& {Brown}}{{Marchetti}
  et~al.}{2018}]{Marchetti}
{Marchetti} T.,  {Rossi} E.~M.,   {Brown} A.~G.~A.,  2018, \mn@doi [\mnras]
  {10.1093/mnras/sty2592}, \href
  {https://ui.adsabs.harvard.edu/\#abs/2018MNRAS.tmp.2466M} {p.~2466}

\bibitem[\protect\citeauthoryear{McKinney}{McKinney}{2010}]{pandas}
McKinney W.,  2010, in van~der Walt S.,  Millman J.,  eds, Proceedings of the
  9th Python in Science Conference. pp 51 -- 56

\bibitem[\protect\citeauthoryear{{McMillan}}{{McMillan}}{2017}]{McMillan_LSR}
{McMillan} P.~J.,  2017, \mn@doi [\mnras] {10.1093/mnras/stw2759}, \href
  {https://ui.adsabs.harvard.edu/\#abs/2017MNRAS.465...76M} {465, 76}

\bibitem[\protect\citeauthoryear{{Meingast}, {Alves}  \&
  {F{\"u}rnkranz}}{{Meingast} et~al.}{2019}]{Meingast_2019}
{Meingast} S.,  {Alves} J.,   {F{\"u}rnkranz} V.,  2019, in The Gaia Universe.
  p.~37, \mn@doi{10.5281/zenodo.2865866}

\bibitem[\protect\citeauthoryear{{Miyamoto} \& {Nagai}}{{Miyamoto} \&
  {Nagai}}{1975}]{Miyamoto_pot}
{Miyamoto} M.,  {Nagai} R.,  1975, \pasj, \href
  {http://adsabs.harvard.edu/abs/1975PASJ...27..533M} {27, 533}

\bibitem[\protect\citeauthoryear{{Myeong}, {Jerjen}, {Mackey}  \& {Da
  Costa}}{{Myeong} et~al.}{2017}]{Myeong_2017}
{Myeong} G.~C.,  {Jerjen} H.,  {Mackey} D.,   {Da Costa} G.~S.,  2017, \mn@doi
  [\apj] {10.3847/2041-8213/aa6fb4}, \href
  {https://ui.adsabs.harvard.edu/\#abs/2017ApJ...840L..25M} {840, L25}

\bibitem[\protect\citeauthoryear{{Myeong}, {Evans}, {Belokurov}, {Amorisco}  \&
  {Koposov}}{{Myeong} et~al.}{2018a}]{Myeong_1}
{Myeong} G.~C.,  {Evans} N.~W.,  {Belokurov} V.,  {Amorisco} N.~C.,   {Koposov}
  S.~E.,  2018a, \mn@doi [\mnras] {10.1093/mnras/stx3262}, \href
  {https://ui.adsabs.harvard.edu/abs/2018MNRAS.475.1537M} {475, 1537}

\bibitem[\protect\citeauthoryear{{Myeong}, {Evans}, {Belokurov}, {Sand ers}  \&
  {Koposov}}{{Myeong} et~al.}{2018b}]{Myeong_2}
{Myeong} G.~C.,  {Evans} N.~W.,  {Belokurov} V.,  {Sand ers} J.~L.,   {Koposov}
  S.~E.,  2018b, \mn@doi [\mnras] {10.1093/mnras/sty1403}, \href
  {https://ui.adsabs.harvard.edu/abs/2018MNRAS.478.5449M} {478, 5449}

\bibitem[\protect\citeauthoryear{{Navarro}, {Frenk}  \& {White}}{{Navarro}
  et~al.}{1996}]{Navarro_Potenial}
{Navarro} J.~F.,  {Frenk} C.~S.,   {White} S. D.~M.,  1996, \mn@doi [\apj]
  {10.1086/177173}, \href
  {https://ui.adsabs.harvard.edu/\#abs/1996ApJ...462..563N} {462, 563}

\bibitem[\protect\citeauthoryear{Newberg, Yanny  \& Willett}{Newberg
  et~al.}{2009}]{Newberg_2009}
Newberg H.~J.,  Yanny B.,   Willett B.~A.,  2009, \mn@doi [The Astrophysical
  Journal] {10.1088/0004-637x/700/2/l61}, 700, L61

\bibitem[\protect\citeauthoryear{{Nissen} \& {Schuster}}{{Nissen} \&
  {Schuster}}{2010}]{Nissen_2010}
{Nissen} P.~E.,  {Schuster} W.~J.,  2010, \mn@doi [\aap]
  {10.1051/0004-6361/200913877}, \href
  {https://ui.adsabs.harvard.edu/\#abs/2010A&A...511L..10N} {511, L10}

\bibitem[\protect\citeauthoryear{{Odenkirchen} et~al.,}{{Odenkirchen}
  et~al.}{2001}]{Odenkirchen_2001}
{Odenkirchen} M.,  et~al., 2001, \mn@doi [\apjl] {10.1086/319095}, \href
  {https://ui.adsabs.harvard.edu/abs/2001ApJ...548L.165O} {548, L165}

\bibitem[\protect\citeauthoryear{Pedregosa et~al.,}{Pedregosa
  et~al.}{2011}]{scikit-learn}
Pedregosa F.,  et~al., 2011, Journal of Machine Learning Research, 12, 2825

\bibitem[\protect\citeauthoryear{Price-Whelan}{Price-Whelan}{2017}]{Gala}
Price-Whelan A.~M.,  2017, \mn@doi [The Journal of Open Source Software]
  {10.21105/joss.00388}, 2, 388

\bibitem[\protect\citeauthoryear{{Price-Whelan} \& {Bonaca}}{{Price-Whelan} \&
  {Bonaca}}{2018}]{2018_example_price_wheel}
{Price-Whelan} A.~M.,  {Bonaca} A.,  2018, \mn@doi [\apj]
  {10.3847/2041-8213/aad7b5}, \href
  {https://ui.adsabs.harvard.edu/abs/2018ApJ...863L..20P} {863, L20}

\bibitem[\protect\citeauthoryear{{Rockosi} et~al.,}{{Rockosi}
  et~al.}{2002}]{Rockosi_2002}
{Rockosi} C.~M.,  et~al., 2002, \mn@doi [\aj] {10.1086/340957}, \href
  {http://adsabs.harvard.edu/abs/2002AJ....124..349R} {124, 349}

\bibitem[\protect\citeauthoryear{{Schlegel}, {Finkbeiner}  \&
  {Davis}}{{Schlegel} et~al.}{1998}]{Reddening}
{Schlegel} D.~J.,  {Finkbeiner} D.~P.,   {Davis} M.,  1998, \mn@doi [\apj]
  {10.1086/305772}, \href
  {https://ui.adsabs.harvard.edu/abs/1998ApJ...500..525S} {500, 525}

\bibitem[\protect\citeauthoryear{{Sch{\"o}nrich}, {Binney}  \&
  {Dehnen}}{{Sch{\"o}nrich} et~al.}{2010}]{Schonrich_2010}
{Sch{\"o}nrich} R.,  {Binney} J.,   {Dehnen} W.,  2010, \mn@doi [\mnras]
  {10.1111/j.1365-2966.2010.16253.x}, \href
  {https://ui.adsabs.harvard.edu/\#abs/2010MNRAS.403.1829S} {403, 1829}

\bibitem[\protect\citeauthoryear{{Sharma}, {Bland-Hawthorn}, {Johnston}  \&
  {Binney}}{{Sharma} et~al.}{2011}]{Sharma_Galaxia}
{Sharma} S.,  {Bland-Hawthorn} J.,  {Johnston} K.~V.,   {Binney} J.,  2011,
  \mn@doi [\apj] {10.1088/0004-637X/730/1/3}, \href
  {https://ui.adsabs.harvard.edu/\#abs/2011ApJ...730....3S} {730, 3}

\bibitem[\protect\citeauthoryear{{Shipp} et~al.,}{{Shipp}
  et~al.}{2018}]{Shipp_2018}
{Shipp} N.,  et~al., 2018, \mn@doi [\apj] {10.3847/1538-4357/aacdab}, \href
  {https://ui.adsabs.harvard.edu/\#abs/2018ApJ...862..114S} {862, 114}

\bibitem[\protect\citeauthoryear{{Taylor}}{{Taylor}}{2005}]{TOPCAT}
{Taylor} M.~B.,  2005, in {Shopbell} P.,  {Britton} M.,   {Ebert} R.,  eds,
  Astronomical Society of the Pacific Conference Series Vol. 347, Astronomical
  Data Analysis Software and Systems XIV. p.~29

\bibitem[\protect\citeauthoryear{{Waskom} et~al.,}{{Waskom}
  et~al.}{2016}]{Seaborn}
{Waskom} M.,  et~al., 2016, {Seaborn: V0.7.1 (June 2016)},
  \mn@doi{10.5281/zenodo.54844}

\bibitem[\protect\citeauthoryear{{Williams} et~al.,}{{Williams}
  et~al.}{2011}]{Williams_Galaxia}
{Williams} M.~E.~K.,  et~al., 2011, \mn@doi [\apj]
  {10.1088/0004-637X/728/2/102}, \href
  {https://ui.adsabs.harvard.edu/\#abs/2011ApJ...728..102W} {728, 102}

\bibitem[\protect\citeauthoryear{{Zinn}, {Pinsonneault}, {Huber}  \&
  {Stello}}{{Zinn} et~al.}{2019}]{Zinn_Paralaxx_Offset}
{Zinn} J.~C.,  {Pinsonneault} M.~H.,  {Huber} D.,   {Stello} D.,  2019, \mn@doi
  [\apj] {10.3847/1538-4357/ab1f66}, \href
  {https://ui.adsabs.harvard.edu/abs/2019ApJ...878..136Z} {878, 136}

\bibitem[\protect\citeauthoryear{{van der Walt}, {Colbert}  \&
  {Varoquaux}}{{van der Walt} et~al.}{2011}]{numpy}
{van der Walt} S.,  {Colbert} S.~C.,   {Varoquaux} G.,  2011, \mn@doi
  [Computing in Science and Engineering] {10.1109/MCSE.2011.37}, \href
  {https://ui.adsabs.harvard.edu/abs/2011CSE....13b..22V} {13, 22}

\makeatother
\end{thebibliography}




\appendix
\section{Orbits and Colour-Magnitude Diagrams for the Replica Data Set}
\label{app:replica-cmds}

\begin{figure*}
    \centering
    \includegraphics[scale=0.2]{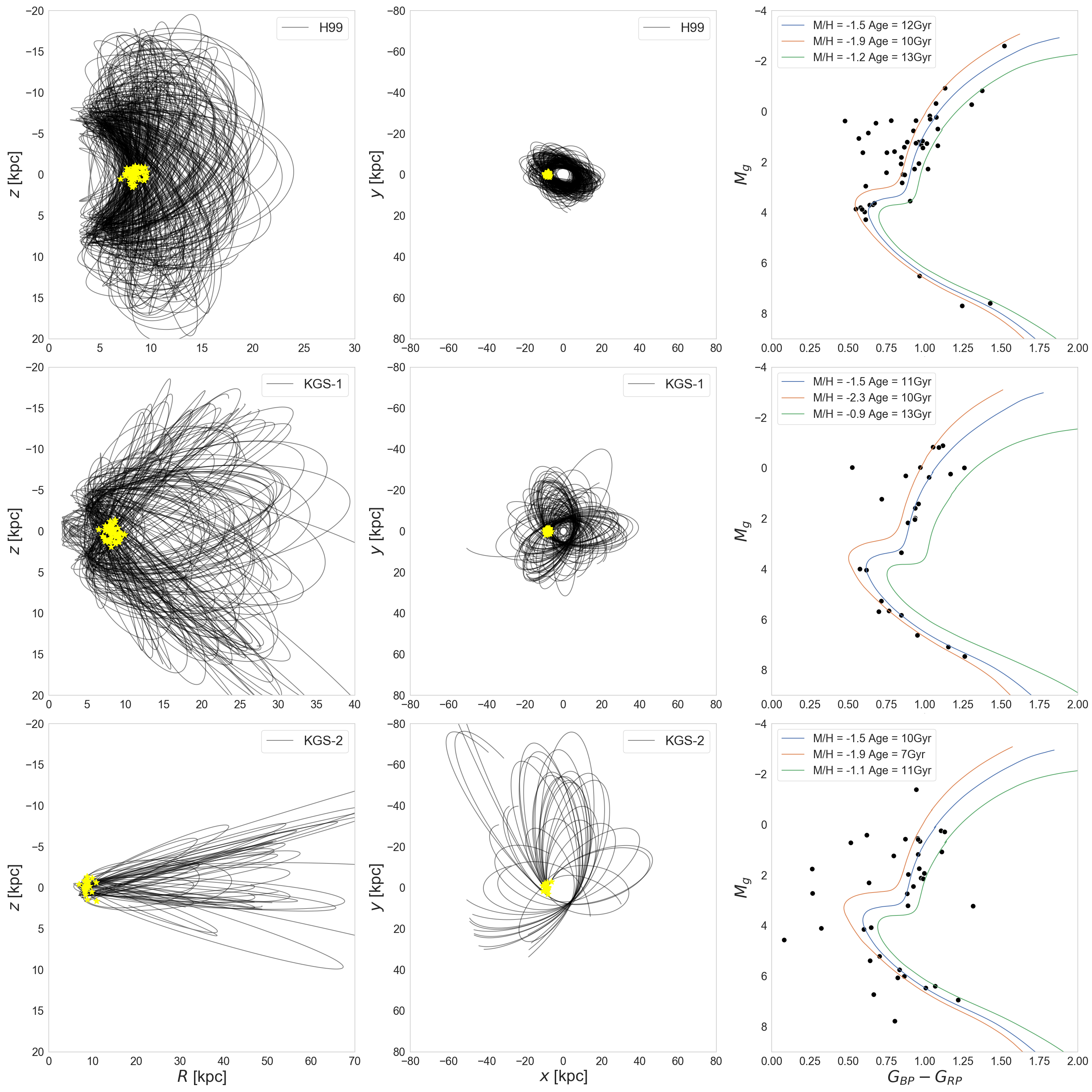}
    \caption{Orbits and photometry for H99, KGS-1 and KGS-2 for the replica data set}
    \label{fig:Replica_rediscovery}
\end{figure*}

\begin{figure*}
    \centering
    \includegraphics[scale=0.2]{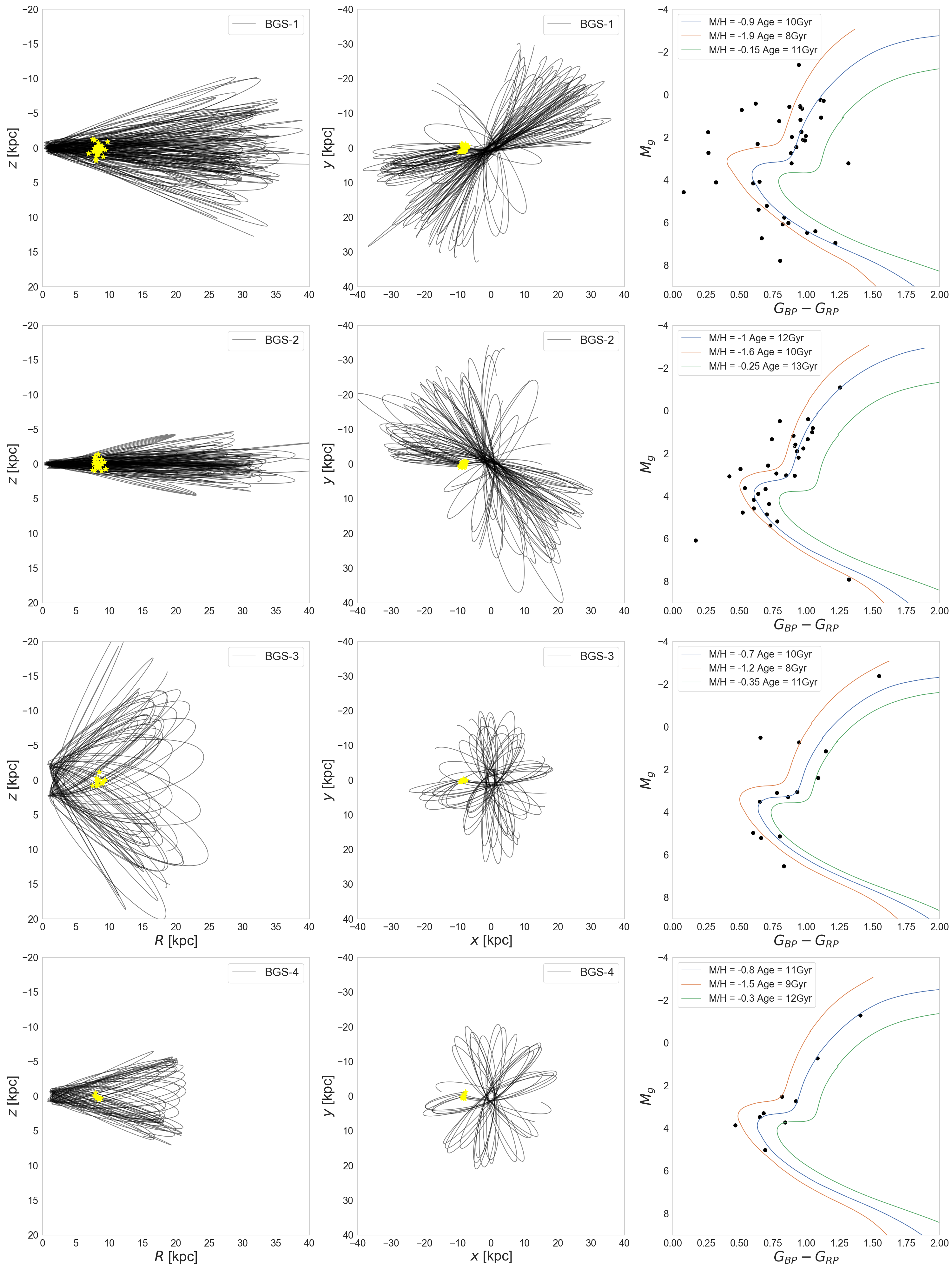}
    \caption{Orbits and photometry for BGS-1, -2, -3 and -4 streams obtained from the replica data set}
    \label{fig:Replica_discovery}
\end{figure*}



\bsp	
\label{lastpage}
\end{document}